\documentclass{article}
\usepackage{slashed,verbatim,subfigure}
\usepackage[numbers,sort&compress]{natbib}
\usepackage{amsmath}
\usepackage{amssymb}
\usepackage{appendix}
\usepackage{graphicx}
\usepackage{dcolumn}
\usepackage{bm}
\usepackage[colorlinks=true,linktocpage=true,
linkcolor=blue,citecolor=blue]{hyperref}
\usepackage[a4paper]{geometry}
\newcommand{\be}{\begin{eqnarray}}
\newcommand{\ee}{\end{eqnarray}}
\newcommand{\ec}{\sigma_{\rm el}}
\newcommand{\dex}{f^{\rm ex}_{\rm aniso}}
\newcommand{\db}{f^{\rm B}_{\rm aniso}}
\newcommand{\df}{f_{\rm iso}}
\begin{document}
\large
\title{\bf{Revisit to electrical and thermal conductivities, Lorenz 
and Knudsen numbers in thermal QCD in a strong magnetic field}}
\author{Shubhalaxmi Rath\footnote{srath@ph.iitr.ac.in}~~and~~Binoy Krishna 
Patra\footnote{binoyfph@iitr.ac.in}\vspace{0.03in} \\ 
Department of Physics, Indian Institute of Technology Roorkee, Roorkee 247667, India}
\date{}
\maketitle
\begin{abstract}
We have explored how the electrical 
($\sigma_{\rm el}$) and thermal ($\kappa$) conductivities in a 
thermal QCD medium get affected in weak-momentum anisotropies
arising either due to a strong magnetic field or due to
asymptotic expansion in a particular direction. This study, in turn, 
facilitates to understand the longevity 
of strong magnetic field through $\ec$, Lorenz number in Wiedemann-Franz 
law, and the validity of
local equilibrium by the Knudsen number through $\kappa$. 
We calculate the conductivities by solving the 
relativistic Boltzmann transport equation in relaxation-time 
approximation, where the interactions are incorporated through
the distribution function within the quasiparticle approach at finite 
$T$ and strong $B$. However, we also compare with the noninteracting 
scenario, which gives unusually large values, thus validating the 
quasiparticle description. We have found that both $\ec$ and $\kappa$ 
get enhanced in a magnetic field-driven anisotropy, but $\ec$ monotonically 
decreases with 
the temperature, opposite to the faster increase in the expansion-driven 
anisotropy. Whereas $\kappa$ increases very slowly with the temperature, 
contrary to its rapid increase in the expansion-driven anisotropy. 
Therefore the conductivities may distinguish the origin of 
anisotropies. The above findings are broadly 
attributed to three factors: the stretching and squeezing of the 
distribution function due to the momentum anisotropies 
generated by the strong magnetic field and asymptotic expansion, 
respectively, the dispersion relation and the
resulting phase-space factor, the relaxation-time in the absence and 
presence of strong magnetic field. Thus $\sigma_{\rm el}$ extracts the 
time-dependence of initially produced strong magnetic field, which 
expectedly decays slower than in vacuum but the expansion-driven 
anisotropy makes the decay faster. The variation in $\kappa$ transpires 
that the Knudsen number ($\Omega$) decreases with the temperature, but the 
expansion-driven anisotropy reduces its magnitude, and the strong magnetic 
field-driven anisotropy raises its value but to less than one, thus the system 
can still be in local equilibrium in a range of temperature and magnetic field. 
Finally, the ratio, $\kappa/\ec$ in Wiedemann-Franz law in magnetic 
field-driven anisotropy increases linearly with temperature but its magnitude 
is smaller than in expansion-driven anisotropic medium. Thus the slope, 
{\em i.e.} the Lorenz number can make the distinction between the anisotropies 
of different origins.

\end{abstract}


\newpage

\section{Introduction}
Relativistic heavy-ion experiments at RHIC and LHC create a new state 
of strongly interacting medium, known as quark gluon plasma (QGP) and 
are continuing to successfully collect the evidences in the form of 
dilepton and photon spectra~\cite{Feinberg:NCA34'1976,Shuryak:PLB78'1978,Kapusta:PRD44'1991}, 
anomalous quarkonium suppression~\cite{Blaizot:PRL77'1996,Satz:NPA783'2007,Rapp:PPNP65'2010}, elliptic flow~\cite{Bhalerao:PLB641'2006,Voloshin:PLB659'2008}, 
jet quenching~\cite{Wang:PRL68'1992,Adcox:PRL88'2002,Chatrchyan:PRC84'2011} etc. for the existence of QGP. The abovementioned predictions were made for the 
simplest possible phenomenological setting, {\em i.e.} 
fully central collisions, where the baryon number density 
is negligible and it is expected that due to the symmetric 
configuration of the collision, no strong magnetic 
fields are produced. But only a small portion of heavy-ion 
collisions  are truly head-on, most collisions indeed occur 
with a finite impact parameter or centrality. As a result, 
the two highly charged ions impacting with a small 
offset may produce extremely large magnetic fields reaching between 
$m_{\pi}^2$ ($\simeq 10^{18}$ Gauss) at RHIC to 15 $m_{\pi}^2$ at 
LHC \cite{Skokov:IJMPA24'2009}.

However, the naive (classical) estimates for the lifetime of 
these strong magnetic fields show that they only exist for a 
small fraction of the lifetime of QGP \cite{QGP}. However, the charge transport 
properties of QGP may significantly extend their 
lifetime, thus the study of the transport coefficients, mainly,
the electrical conductivity ($\sigma_{\rm el}$) becomes essential.  
Here our motivations are twofold, which complement each 
other: First, we wish to revisit $\sigma_{\rm el}$ in an isotropic hot 
QCD medium to check how long the magnetic 
field produced in relativistic heavy ion collision stays appreciably
large, {\em i.e.}, some sort of time-dependence of nascent 
magnetic field. However, the issue about the longevity of the magnetic 
field is not yet settled. So keeping the uncertainties about 
the exact nature of magnetic field in mind, if the 
external magnetic field still remains large, the transport properties 
of the medium can be significantly affected by the strong magnetic 
field. Our second motivation is to quantify the 
effect on electrical and thermal ($\kappa$) conductivities. Since 
$\sigma_{\rm{el}}$ is responsible for the production of electric 
current due to the Lenz's law, its value becomes vital for the strength 
of chiral magnetic effect~\cite{Fukushima:PRD78'2008}. Moreover the 
electric field in mass asymmetric collisions has overall a preferred 
direction, which will eventually generate a charge asymmetric flow and 
the strength of the flow  is given by 
$\sigma_{\rm{el}}$~\cite{Hirono:PRC90'2014}. 
Furthermore, $\sigma_{\rm{el}}$ is used as a vital input for many
phenomenological applications in RHIC, LHC etc., such as
the emission rate of soft photons~\cite{Kapusta:BOOK'2006}.
The effects of magnetic fields on $\sigma_{\rm el}$ for quark 
matter have been investigated previously in different models, such 
as quenched SU($2$) lattice gauge theory 
\cite{Buividovich:PRL105'2010}, the dilute instanton-liquid 
model \cite{Nam:PRD86'2012}, the nonlinear electromagnetic currents 
\cite{Kharzeev:PPNP75'2014,Satow:PRD90'2014}, axial Hall current 
\cite{Pu:PRD91'2015}, real-time formalism using the diagrammatic method 
\cite{Hattori:PRD94'2016}, effective fugacity approach~\cite{Kurian:PRD96'2017} 
etc. 

In ultrarelativistic heavy ion collisions, it is observed that the 
suppression of charged particle production gets reduced while going 
from central to noncentral Pb$+$Pb collisions 
\cite{Aamodt:PLB696'2011,Foka:RP1'2016}, where strong magnetic field 
exists too. Therefore such strong magnetic field might significantly 
affect the production of particles especially the quarks which 
are produced within short time scale less than $\sim 1$ fm and 
can alter their dynamics too. A comprehensive studies on how the 
particle production gets influenced by an external strong magnetic field 
are nicely depicted in refs. \cite{Tuchin:PRD91'2015,Basar:PRL109'2012,
Fukushima:PRD86'2012}. The motion of quark in strong magnetic
field becomes effectively one dimensional, which in turn enhances the 
quark-antiquark attraction and makes it favorable for pair production
and the quark pairs get polarized along the direction of magnetic field 
\cite{Kabat:PRD66'2002}. As a result, one might expect that the magnetic
field might affect the anisotropic flow of the particles.

As we know already that the external magnetic field 
modifies the dispersion relation of a charged particle 
($\omega_{i,n} = \sqrt{p_L^2 +2n|q_iB|+m_i^2}$) 
quantum mechanically, where the motion along the 
longitudinal direction ($p_L$) (with respect to the magnetic field 
direction) remains the same as for a free particle and 
only the motion along the transverse direction ($p_T$) gets 
quantized in terms of the Landau levels ($n$). In strong magnetic field (SMF) 
limit ($eB \gg T^2$ as well as $eB \gg m_i^2$), only the lowest Landau 
level (LLL) will be occupied, {\em i.e.} $p_T \approx 0$, and the particle can 
only move along the direction of the magnetic field, resulting 
an anisotropy in the momentum space, {\em i.e.} $p_L \gg p_T$.
Thus the anisotropic parameter, $\xi$ ($\frac{\langle {\bf p}_{T}^{2}
\rangle}{2\langle p_{L}^{2}\rangle}-1$) comes out to be negative and 
for a weak-anisotropy ($\xi <1$), the distribution function may be 
approximated by stretching the isotropic one along a certain direction 
(say, the direction of magnetic field). Thus, to know the effects of strong 
magnetic field on conductivities in kinetic theory approach, an 
introduction of anisotropy is automatically needed.

Much earlier than the former one, it was envisaged 
that the relativistic heavy ion collisions at 
the initial stage may induce a momentum anisotropy in the 
local rest frame of fireball, due to the asymptotic free expansion
of the fireball in the beam direction compared to its transverse 
direction~\cite{Dumitru:PLB662'2008,Dumitru:PRD79'2009}. Unlike the 
previous one, here, $p_T$ is greater than $p_L$, hence the anisotropy
parameter becomes positive. Therefore, 
for a weak-anisotropy ($\xi <1$), the distribution of partons
can be approximated by squeezing the isotropic one along a certain 
direction and its effects on many phenomenological and theoretical 
observations have already been made. {\em For example}, the leading-order 
dilepton and photon yields get enhanced due to the anisotropic component 
\cite{Martinez:PRC78'2008, Ryblewski:PRD92'2015, Mukherjee:EPJA53'2017,
Bhattacharya:PRD93'2016}. Recently, one of us had observed 
the effect of this kind of anisotropy on the properties 
of heavy quarkonium bound states~\cite{Thakur:PRD88'2013} and the electrical 
conductivity \cite{Srivastava:PRC91'2015}, where the heavy quarkonia 
are found to dissociate earlier than its counterpart in isotropic one 
and the electrical conductivity decreases with the 
increase of anisotropy. Later its relation with the shear viscosity was 
explored in \cite{Thakur:PRD95'2017}. Besides the abovementioned 
anisotropies, the event-by-event fluctuations of heavy ion collisions also 
produce the anisotropy, which plays a crucial role in understanding new phenomena 
such as the elliptic flow, the triangular flow \cite{Bhalerao:PRC84'2011,Heinz:PRC87'2013} etc. In fact, the anisotropy 
generated by the event-by-event fluctuations also transpires to the final 
anisotropic flow angles.

Now we move on to the thermal conductivity ($\kappa$), which is 
related to the efficiency of the heat flow or the energy 
dissipation in a thermal QCD medium. Our intention is to 
comment on the range of temperature and possibly, magnetic field, 
in which the assumption of local equilibrium in hydrodynamics can be 
validated in terms of Knudsen number ($\Omega$). The Knudsen number 
is the ratio of the mean free path ($\lambda$) to the 
characteristic length of the system, where  $\lambda$ in turn is related 
to $\kappa$ ($\lambda=3\kappa/(vC_V)$). Similar to the 
electrical conductivity, we also wish to explore the effect of strong 
magnetic field on the 
thermal conductivity by calculating it in the presence of 
weak-momentum anisotropy caused by the strong magnetic field. 
A natural question arises about whether we can distinguish the 
anisotropies through the transport coefficients. Knowing that, we can 
improve the knowledge on the transport properties of the medium. This 
query might be worthy of investigation.

The electronic contribution of the thermal and electrical 
conductivities are not completely independent, {\em rather} their ratio is 
equal to the product of Lorenz number ($L$) and temperature, widely known 
as Wiedemann-Franz law. In fact, the ratio, $\kappa/\sigma_{\rm el}$ has 
approximately the same value for different metals at the same temperature. 
But, it diverges in quasi-one-dimensional metallic phase with 
decreasing temperature~\cite{Mahajan:PRB88'2013}, reaching a value much 
larger than that found in conventional metals nearer to the insulator-metal 
transition~\cite{Proust:PRB72'2005}, thermally populated electron-hole plasma 
in graphene \cite{Crossno:S351'2016} etc. Recently, the temperature 
dependence of the Lorenz number was calculated for the two-flavor quark 
matter in NJL model \cite{Harutyunyan:PRD95'2017} and 
for the strongly interacting QGP medium \cite{Mitra:PRD96'2017}.
In the metallic phase, the electronic contribution to thermal conductivity 
is much smaller than what would be expected from the Wiedemann-Franz law, 
which can be explained in terms of independent propagation of charges and heat 
in a strongly correlated system. However, in this work we intend to observe 
how the ratio gets affected due to the presence of an ambient strong magnetic 
field, which in turn generates the anisotropy.

In this work, we have evaluated both the conductivities in kinetic theory 
approach, where the relativistic Boltzmann transport equation (RBTE) is 
employed and is solved in the relaxation-time approximation (RTA), 
where, as such, there is no scope to incorporate the interaction among the  partons\footnote{If one can solve RBT equation with the
collisional integral ($C[f]$), one can then incorporate the 
interactions through the matrix element.}. 
We circumvent the problem by incorporating 
the interactions among partons through their dispersion relations,
known as quasiparticle model (QPM), 
in their distribution functions. The quasiparticle masses 
are conveniently obtained from their respective
self energies, which, in turn, depend on the temperature and
the magnetic field. Thus the presence of magnetic field 
affects both electrical and thermal 
conductivities. However, as a base line, we also compute the conductivities 
with the current quark masses (noninteracting), which give unusually large 
values, thus motivating us to use the quasiparticle model.

In brief, we have observed that the 
electrical and thermal conductivities of the hot QCD medium get
enhanced in the presence of strong magnetic field-driven anisotropy, 
compared to the counterparts in the expansion-driven 
anisotropic medium. We have also noticed that the 
unusually large values of conductivities in the noninteracting  scenario
have been circumvented in the quasiparticle model. 
As a corollary, the ratio, $\kappa/\ec$ in a strong 
magnetic field shows linear enhancement with the temperature, whose 
magnitude and slope are larger than in isotropic medium but smaller 
than in expansion-driven anisotropic medium, thus describing 
different Lorenz numbers ($\kappa/(\ec T)$). We have also 
observed that the presence of strong magnetic field makes 
the Knudsen number larger (but remains less than one) than its value in the 
(an)isotropic medium. Therefore the transport coefficients and their ratio 
might help to distinguish the origin of 
aforesaid anisotropies in a thermal medium produced at the initial 
stage of ultrarelativistic heavy ion collision. However, in our 
present attempt, we are not exploring the anisotropy 
produced due to the even-by-event fluctuations.

The present work is organized as follows. In section 2, 
we have first revisited the 
electrical conductivity for an isotropic thermal medium and then proceeded 
for the anisotropic thermal mediums due to expansion-induced 
and strong magnetic field-induced anisotropies with the current quark masses.
Similarly in section 3 we have done the same for the thermal conductivity. The 
Wiedemann-Franz law and the Knudsen number are revisited in section 
4 in light of the observations made in sections 2 and 3. In section 5, we have 
introduced the quasiparticle mass in the presence of both temperature and strong 
magnetic field and recomputed the results on electrical and thermal 
conductivities, which in turn redefined the Wiedemann-Franz law 
and the Knudsen number. Finally, we have concluded our results and
future outlook in section 6.

\section{Electrical conductivity}
Transport coefficients such as electrical conductivity and thermal 
conductivity of a hot QCD system can be determined using different 
models and approaches namely relativistic Boltzmann transport equation \cite{Muronga:PRC76'2007,Puglisi:PRD90'2014,Thakur:PRD95'2017,Yasui:PRD96'2017}, 
the Chapman-Enskog approximation \cite{Mitra:PRD94'2016,Mitra:PRD96'2017}, the 
correlator technique using Green-Kubo formula \cite{Nam:PRD86'2012,Greif:PRD90'2014,Feng:PRD96'2017} and the lattice simulation \cite{Gupta:PLB597'2004,Aarts:JHEP1502'2015,Ding:PRD94'2016}. 
However, we will employ the relativistic Boltzmann transport equation with the 
relaxation-time approximation to calculate the electrical conductivity for both 
isotropic and anisotropic hot QCD mediums in subsections 2.1 and 2.2, respectively.

\subsection{Electrical conductivity for an isotropic thermal medium}
When an isotropic and hot medium of quarks, antiquarks and 
gluons in thermal equilibrium is disturbed infinitesimally 
by an electric field, an electric current $J_\mu$ is induced and is given by
\begin{eqnarray}\label{current}
J_\mu = \sum_i q_i g_i \int\frac{d^3\rm{p}}{(2\pi)^3\omega_i}
p_\mu [\delta f_i^q(x,p)+\delta f_i^{\bar q}(x,p)]
~,\end{eqnarray}
where the summation is over three flavors ($u$, $d$ and $s$) and 
$q_i$, $g_i$ and $\delta f_i^q $ ($\delta f_i^{\bar q}$) are the 
electric charge, degeneracy factor and infinitesimal change in the 
distribution function for the quark (antiquark) of $i$th flavor, 
respectively. In our calculations, we will be using 
$\delta f_i^q=\delta f_i^{\bar q}=\delta f_i$, for zero chemical 
potential. According to Ohm's law, the longitudinal component of the 
spatial part of four-current is directly proportional to the external 
electric field and the proportionality factor is known as the 
electrical conductivity,
\be
\mathbf{J}=\sigma_{\rm el}\mathbf{E}
~.\ee
The infinitesimal change in quark distribution function is defined as 
$\delta f_i=f_i-f_i^{\rm iso}$, where $f_i^{\rm iso}$ is the 
equilibrium distribution function in the isotropic medium for 
$i$th flavor,
\be\label{D.F.}
f_i^{\rm iso}=\frac{1}{e^{\beta\omega_i}+1}
~,\ee
with $\omega_i=\sqrt{\mathbf{p}^2+m_i^2}$. It is possible to 
obtain $\delta f_i$ from the relativistic Boltzmann transport 
equation (RBTE) \cite{Crecignani:2002},
\be\label{R.B.T.E.(1)}
p^\mu\frac{\partial f_i(x,p)}{\partial x^\mu}+q_i F^{\rho\sigma} 
p_\sigma \frac{\partial f_i(x,p)}{\partial p^\rho}=\mathcal{C}[f_i(x,p)]
~,\ee
where $F^{\rho\sigma}$ denotes the electromagnetic 
field strength tensor and the collision term, $\mathcal{C}[f_i(x,p)]$ 
in the relaxation-time approximation is given by
\be
\mathcal{C}[f_i(x,p)] \simeq -\frac{p_\nu u^\nu}{\tau_i}\delta f_i(x,p)
~,\ee
where $u^\nu$ is the four-velocity of fluid in the local rest 
frame and the relaxation-time ($\tau_i$) for $i$th 
flavor in a thermal medium is given \cite{Hosoya:NPB250'1985} by
\be
\tau_i=\frac{1}{5.1T\alpha_s^2\log\left(1/\alpha_s\right)\left[1+0.12(2N_i+1)\right]}
~.\ee

To see the response of electric field, we use 
only $\rho=i$ and $\sigma=0$ and {\em vice versa}, components of the 
electromagnetic field strength tensor, {\em i.e.} 
$F^{i0}=\mathbf{E}$ and $F^{0i}=-\mathbf{E}$ in our calculation, thus the 
RBTE (\ref{R.B.T.E.(1)}) takes the following form,
\be
q_i\mathbf{E}\cdot\mathbf{p}\frac{\partial f_i^{\rm iso}}{\partial p_0}
+q_i p_0\mathbf{E}\cdot\frac{\partial f_i^{\rm iso}}{\partial \mathbf{p}}
=-\frac{p_0}{\tau_i}\delta f_i
~.\ee
Hence the infinitesimal disturbance is obtained as
\be
\delta f_i=2q_i\tau_i\beta\frac{\mathbf{E}\cdot\mathbf{p}}{\omega_i}
f_i^{\rm iso}(1-f_i^{\rm iso})
~.\ee
Now substituting the value of $\delta f_i$ in eq. (\ref{current}), 
we obtain the electrical conductivity for an isotropic thermal medium,
\be\label{I.E.C.}
\sigma_{\rm el}^{\rm iso}=\frac{2\beta}{3\pi^2}\sum_i g_i q_i^2\int d{\rm p}~\frac{{\rm p}^4}{\omega_i^2} ~ \tau_i ~ f_i^{\rm iso}(1-f_i^{\rm iso})
~,\ee
which can now be used to show how the magnetic field varies 
with time in the isotropic thermal conducting medium. 
According to electrodynamics, the magnetic field created in 
vacuum due to the spatial variation of the electric field, 
rapidly changes over time. However for a medium with 
substantial value of electrical conductivity, the momentary 
magnetic field would induce an electric current which ultimately 
would help to enhance the lifetime of the strong magnetic field.

\subsection{Electrical conductivity for an anisotropic thermal medium}
Here we will mainly discuss two types of momentum anisotropies, which may
arise in the very early stages of ultrarelativistic heavy ion collisions.
The first one is due to the preferential flow in the longitudinal direction 
compared to the transverse direction and the second one is due to the
creation of a strong magnetic field. We will first revisit the former one.

\subsubsection{Expansion-induced anisotropy}
At early times, the QGP created in the heavy ion collisions 
experiences larger longitudinal expansion than the radial 
expansion and this develops a local momentum anisotropy. 
For the weak-momentum anisotropy ($\xi<1$) in a particular 
direction (say $\mathbf{n}$), the distribution function can 
be approximated from the isotropic one \cite{Romatschke:PRD68'2003} as
\be\label{A.D.F.}
f_{{\rm ex},i}^{\rm aniso}(\mathbf{p};T)=\frac{1}{e^{\beta\sqrt{\rm{p}^2+\xi(\mathbf{p}\cdot\mathbf{n})^2+m_i^2}}+1}
~,\ee
which can be expanded in Taylor series, and up to 
$\mathcal{O}(\xi)$ it takes the following form,
\be\label{expansion}
f_{{\rm ex},i}^{\rm aniso}=f_i^{\rm iso}-\frac{\xi\beta(\mathbf{p}\cdot\mathbf{n})^2}{2\omega_i}f_i^{\rm iso}(1-f_i^{\rm iso})
~.\ee
The anisotropic parameter ($\xi$) is generically defined in terms 
of the transverse and longitudinal components of momentum as
\be\label{parameter}
\xi=\frac{\left\langle\mathbf{p}_T^2\right\rangle}{2\left\langle p_L^2\right\rangle}-1
~,\ee
where $p_L=\mathbf{p}\cdot\mathbf{n}$, $\mathbf{p}_T=\mathbf{p}-\mathbf{n}\cdot(\mathbf{p}\cdot\mathbf{n})$, $\mathbf{p}\equiv(\rm{p}\sin\theta\cos\phi,\rm{p}\sin\theta\sin\phi,\rm{p}\cos\theta)$, $\mathbf{n}=(\sin\alpha,0,\cos\alpha)$, $\alpha$ is the angle between 
z-axis and direction of anisotropy, $(\mathbf{p}\cdot\mathbf{n})^2=\rm{p}^2c(\alpha,\theta,\phi)=\rm{p}^2(\sin^2\alpha\sin^2\theta\cos^2\phi+\cos^2\alpha\cos^2\theta
+\sin(2\alpha)\sin\theta\cos\theta\cos\phi)$. For $p_T\gg p_L$, $\xi$ takes 
positive value, which explains the removal of particles with a large 
momentum component along the $\mathbf{n}$ direction due to the 
faster longitudinal expansion than the transverse expansion 
\cite{Dumitru:PRD79'2009}.

Now we are going to observe how the weak-momentum anisotropy 
affects the electrical conductivity of the thermal medium. Thus, 
after solving the RBTE (\ref{R.B.T.E.(1)}) for the anisotropic 
distribution function, we get $\delta f_i$ as
\be
\nonumber\delta f_i &=& \frac{2\tau_i \beta q_i\mathbf{E}\cdot\mathbf{p}}{\omega_i}\left[f_i^{\rm iso}(1-f_i^{\rm iso})+\frac{\xi c(\alpha,\theta,\phi)}{2}\left\lbrace -\frac{\beta\rm{p}^2}{\omega_i}f_i^{\rm iso}(1-f_i^{\rm iso})\right.\right. \\ && \left.\left.\hspace{2 cm}+\frac{2\beta\rm{p}^2}{\omega_i}{f_i^{\rm iso}}^2(1-f_i^{\rm iso})-\frac{\rm{p}^2}{\omega_i^2}f_i^{\rm iso}(1-f_i^{\rm iso})+f_i^{\rm iso}(1-f_i^{\rm iso}) \right\rbrace\right]
,\ee
which is then substituted in eq. (\ref{current}) to yield the 
expression of electrical conductivity for an expansion-driven anisotropic medium,
\be\label{A.E.C.}
\nonumber\sigma_{\rm el,ex}^{\rm aniso} &=& \frac{2\beta}{3\pi^2}\sum_i g_i q_i^2\int d{\rm p}\frac{{\rm p}^4}{\omega_i^2} ~ \tau_i ~ f_i^{\rm iso}(1-f_i^{\rm iso})-\frac{\xi\beta^2}{9\pi^2}\sum_i g_i q_i^2\int d{\rm p}\frac{{\rm p}^6}{\omega_i^3} ~ \tau_i ~ f_i^{\rm iso}(1-f_i^{\rm iso}) \\ && \nonumber+\frac{2\xi\beta^2}{9\pi^2}\sum_i g_i q_i^2\int d{\rm p}\frac{{\rm p}^6}{\omega_i^3} ~ \tau_i ~ {f_i^{\rm iso}}^2(1-f_i^{\rm iso})-\frac{\xi\beta}{9\pi^2}\sum_i g_i q_i^2\int d{\rm p}\frac{{\rm p}^6}{\omega_i^4} ~ \tau_i ~ f_i^{\rm iso}(1-f_i^{\rm iso}) \\ && +\frac{\xi\beta}{9\pi^2}\sum_i g_i q_i^2\int d{\rm p}\frac{{\rm p}^4}{\omega_i^2} ~ \tau_i ~ f_i^{\rm iso}(1-f_i^{\rm iso})
~,\ee
where the first term in R.H.S. is the electrical conductivity for an 
isotropic medium. So in terms of $\sigma_{\rm el}^{\rm iso}$, 
$\sigma_{\rm el,ex}^{\rm aniso}$ is written as
\be\label{A.E.C.(1)}
\nonumber\sigma_{\rm el,ex}^{\rm aniso} &=& \sigma_{\rm el}^{\rm iso} - \xi\left[\frac{\beta^2}{9\pi^2}\sum_i g_i q_i^2\int d{\rm p}\frac{{\rm p}^6}{\omega_i^3}~ \tau_i ~ f_i^{\rm iso}(1-f_i^{\rm iso})\left\lbrace 1-2f_i^{\rm iso}+\frac{1}{\beta\omega_i} \right\rbrace\right. \\ && \left.-\frac{\beta}{9\pi^2}\sum_i g_i q_i^2\int d{\rm p}\frac{{\rm p}^4}{\omega_i^2}~ \tau_i ~ f_i^{\rm iso}(1-f_i^{\rm iso})\right]
.\ee

\subsubsection{Life-span of magnetic field}
Earlier, people had thought that the magnetic field generated 
in the heavy ion collision decays instantly. However in the 
presence of transport coefficient such as electrical 
conductivity, the lifetime of magnetic field may 
be elongated. To reaffirm this, we are going to see the 
variation of magnetic field using the value of electrical 
conductivity that we have calculated above for both 
isotropic and anisotropic mediums.

\begin{figure}[]
\begin{center}
\begin{tabular}{c c}
\includegraphics[width=7.4cm]{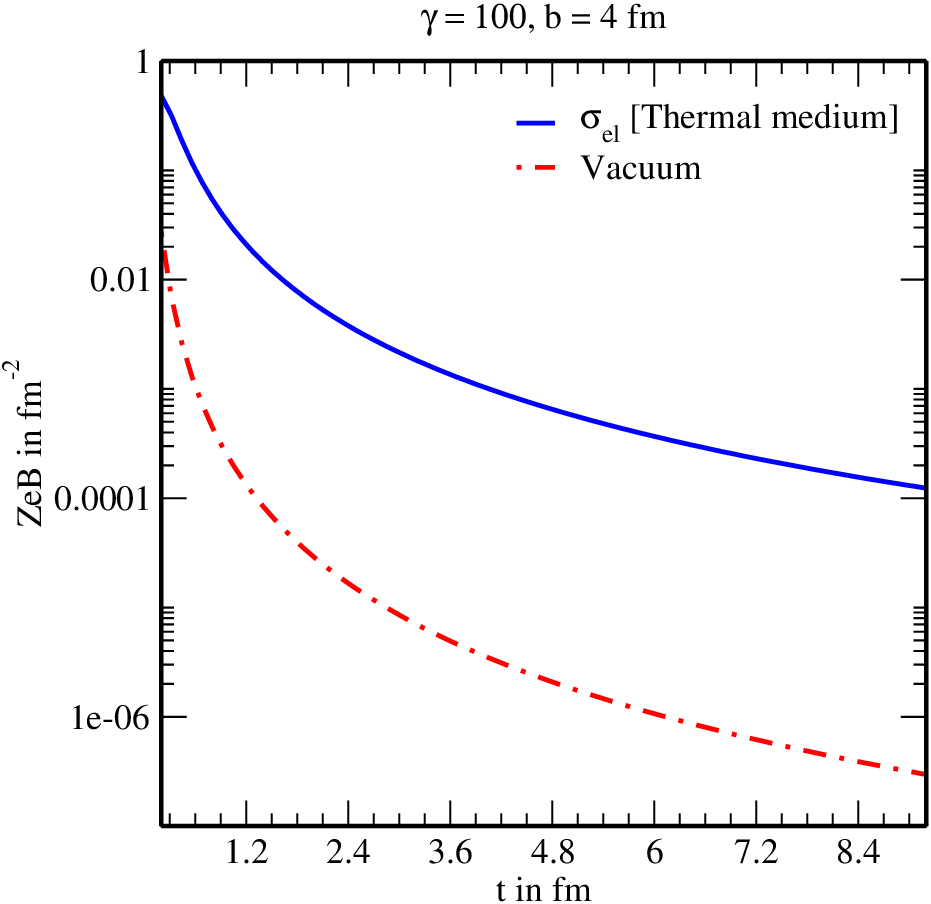}&
\hspace{0.3 cm}
\includegraphics[width=7.4cm]{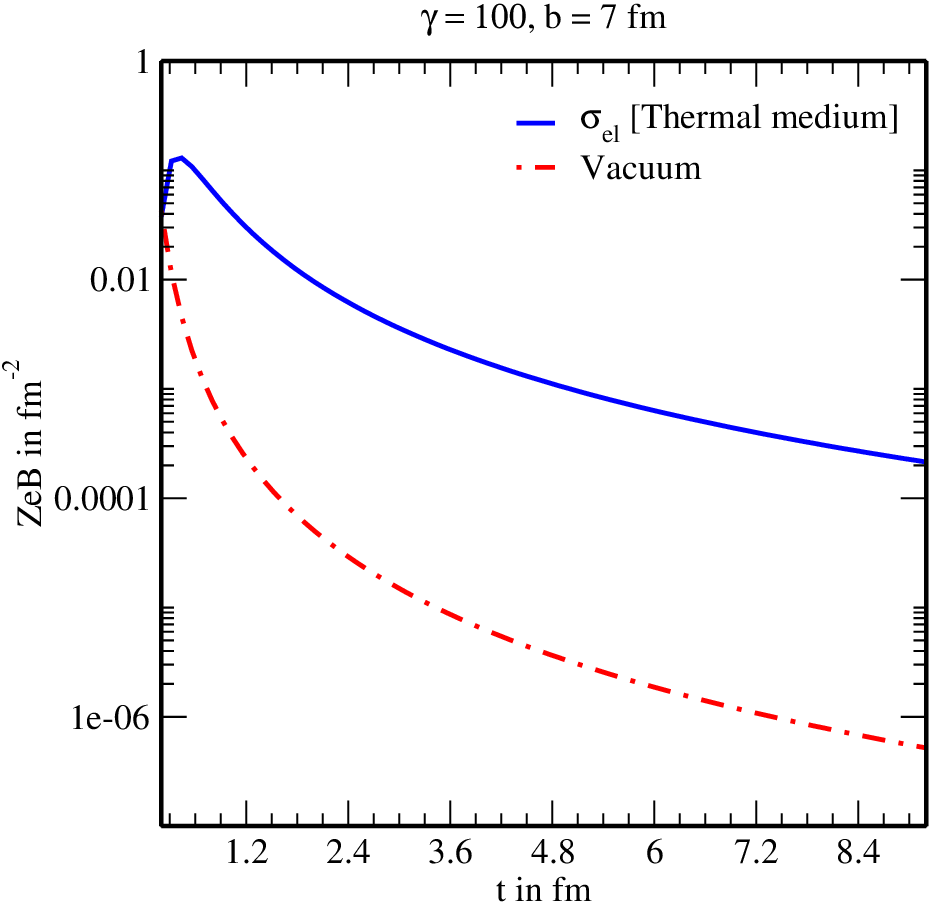} \\
a & b
\end{tabular}
\caption{Comparison between the variations of magnetic field with time 
in an isotropic thermal conducting medium and in a vacuum for two values of the 
impact parameter (a) $b=4$ fm and (b) $b=7$ fm, with the Lorentz 
factor $\gamma=100$ for Au$+$Au collision at RHIC energy 
$\sqrt{s}=200$ GeV and $Z=79$ (gold nucleus).}\label{eb.iso1}
\end{center}
\end{figure}

\begin{figure}[]
\begin{center}
\begin{tabular}{c c}
\includegraphics[width=7.4cm]{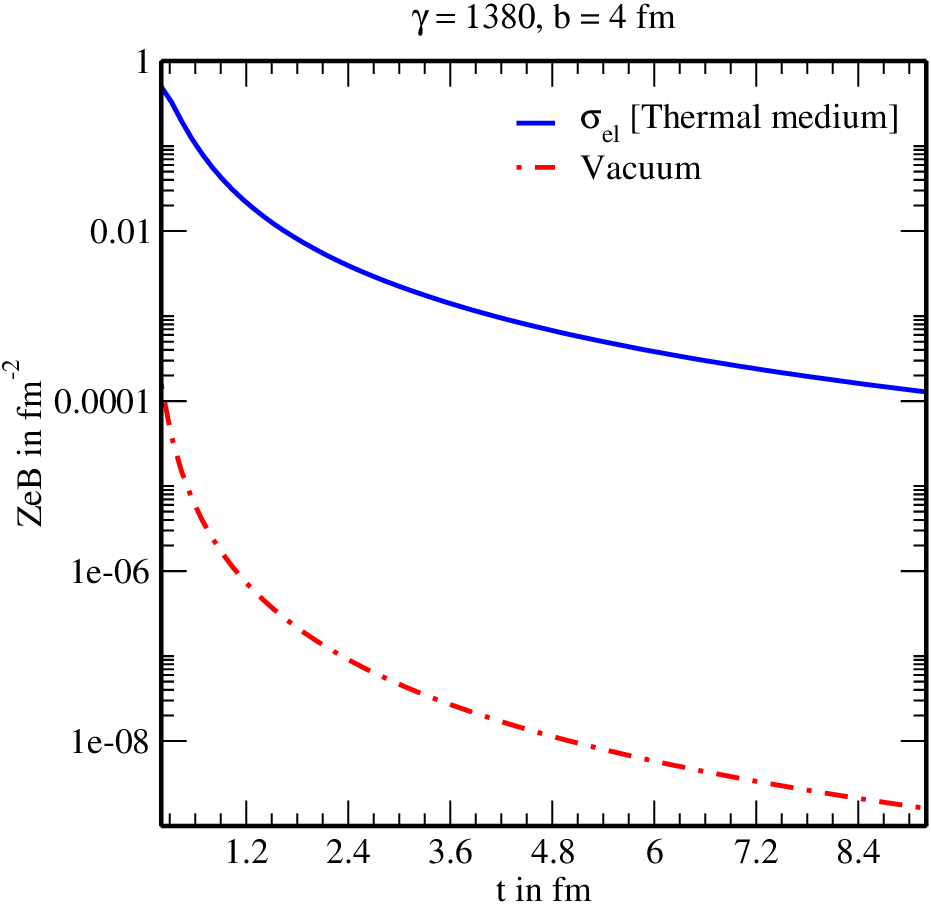}&
\hspace{0.3 cm}
\includegraphics[width=7.4cm]{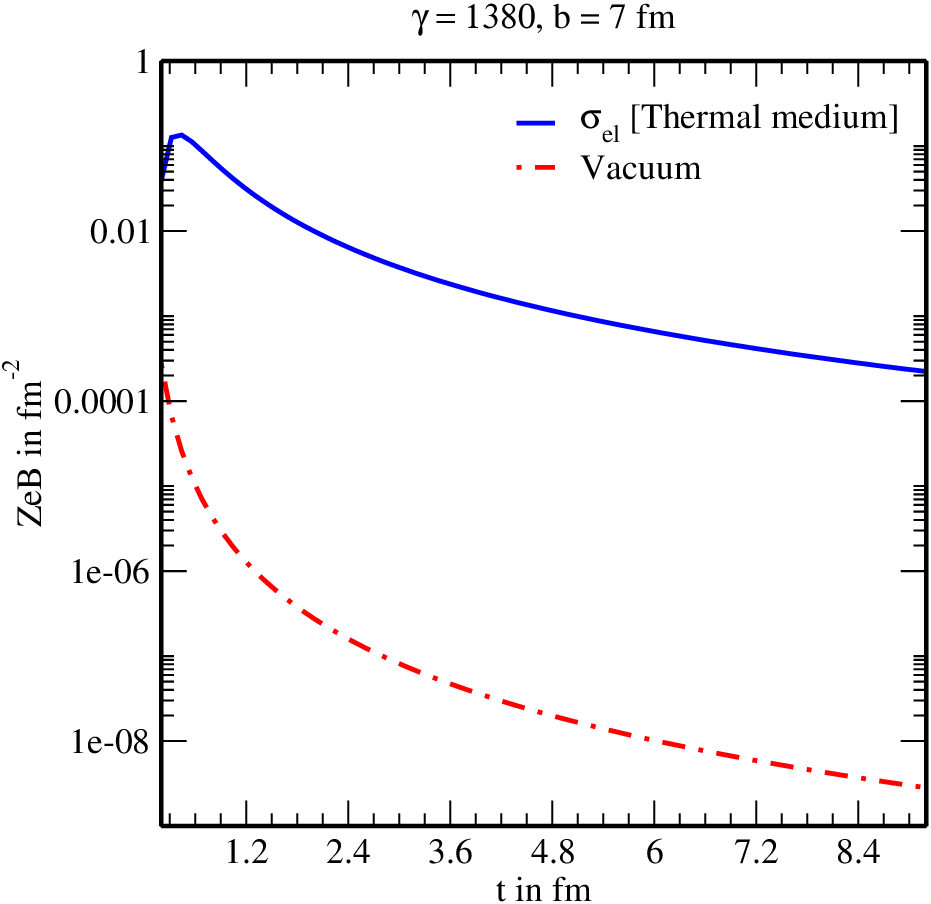} \\
a & b
\end{tabular}
\caption{Comparison between the variations of magnetic field with time 
in an isotropic thermal conducting medium and in a vacuum for two values of the 
impact parameter (a) $b=4$ fm and (b) $b=7$ fm, with the Lorentz 
factor $\gamma=1380$ for Pb$+$Pb collision at LHC energy 
$\sqrt{s}=2.76$ TeV and $Z=82$ (lead nucleus).}\label{eb.iso2}
\end{center}
\end{figure}

Thus, for a charged particle moving in $x$-direction, a 
magnetic field will be produced in the perpendicular 
direction of the particle trajectory, say $z$-direction. 
According to the Maxwell's equations, the magnetic field 
created along $z$-direction is expressed, as a function 
of time and electrical conductivity 
\cite{Tuchin:AHEP2013'2013} for an isotropic medium as
\be\label{eb1}
e\mathbf{B}^{\rm iso}_{\rm medium}=\frac{e^2b\sigma_{\rm el}^{\rm iso}}{8\pi(t-x)^2}e^{-\frac{b^2\sigma_{\rm el}^{\rm iso}}{4(t-x)}}\hat{\mathbf{z}}
~.\ee
However for an anisotropic medium, the expression for 
$e\mathbf{B}$ is not available, so we assumed the same 
expression by replacing 
$\sigma_{\rm el}^{\rm iso}\rightarrow\sigma_{\rm el,ex}^{\rm aniso}$, at 
least, for weak-anisotropy,
\be\label{eb2}
e\mathbf{B}^{\rm aniso}_{\rm medium}=\frac{e^2b\sigma_{\rm el,ex}^{\rm aniso}}{8\pi(t-x)^2}e^{-\frac{b^2\sigma_{\rm el,ex}^{\rm aniso}}{4(t-x)}}\hat{\mathbf{z}}
~.\ee
For the sake of comparison, the magnetic field produced in vacuum  \cite{Tuchin:AHEP2013'2013} is given by,
\be\label{ebv}
e\mathbf{B}_{\rm vacuum}=\frac{e^2b\gamma}{4\pi\left\lbrace b^2+\gamma^2(t-x)^2 \right\rbrace^{3/2}}\hat{\mathbf{z}}
~,\ee
where $b$ and $\gamma$ denote the impact parameter and the 
Lorentz factor of heavy ion collision, respectively. In 
equations (\ref{eb1}) and (\ref{eb2}), the electrical 
conductivity is taken as a function of time through the 
cooling law, $T^3\propto{t^{-1}}$, where the initial time 
and the temperature are set at $0.2$ fm and $390$ MeV, 
respectively. From figures \ref{eb.iso1} and \ref{eb.iso2}, 
which are plotted at $x=0$ for the centre of mass energies 
$200$ GeV and $2.76$ TeV, respectively, we see that the 
magnetic field in the isotropic conducting medium decays 
very slowly as compared to the vacuum. At initial time, the 
fluctuation of magnetic field in a thermal medium is quite 
high, however after certain time, it gradually stabilizes.

\begin{figure}[]
\begin{center}
\includegraphics[width=7.4cm]{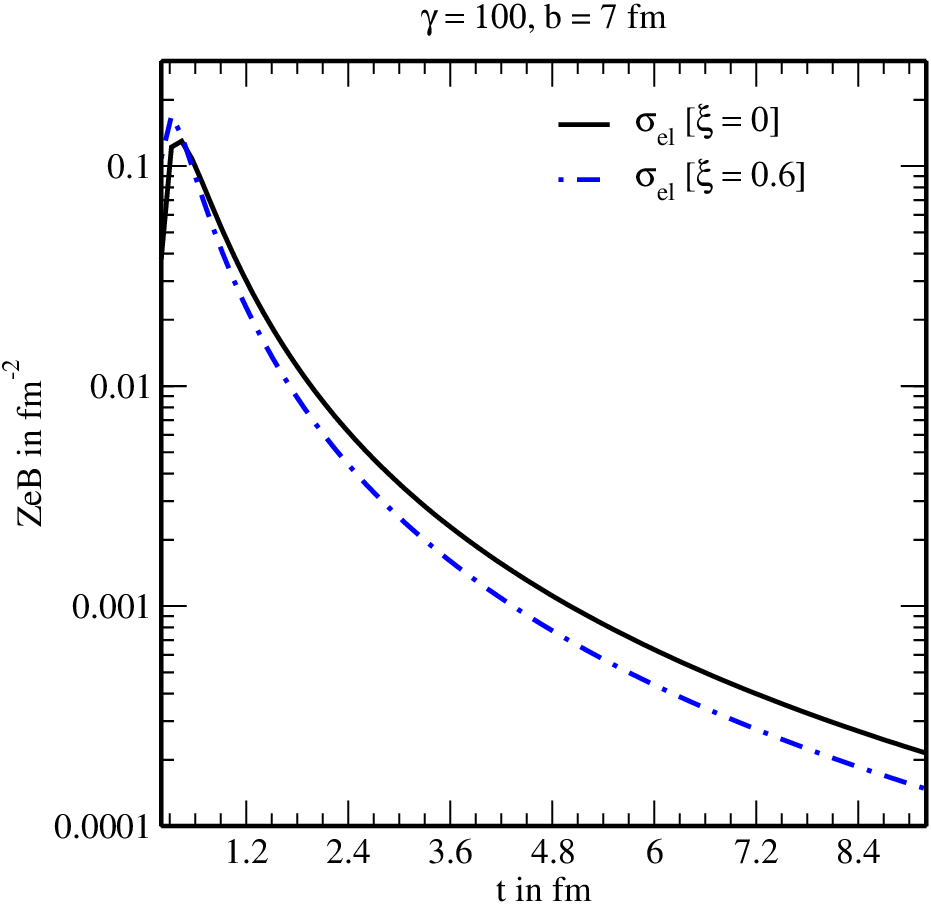}
\caption{Comparison between the variations of magnetic field with time 
in isotropic and anisotropic thermal conducting mediums for impact 
parameter $b=7$ fm, Lorentz factor $\gamma=100$ and 
$Z=79$ (gold nucleus).}\label{eb.aniso}
\end{center}
\end{figure}

However, for a conducting medium in the presence of 
weak-momentum anisotropy ($\xi=0.6$), we have 
observed (from figure \ref{eb.aniso}) that 
the lifetime of existence of a nearly stable magnetic 
field in the anisotropic thermal medium is slightly less 
than in the isotropic thermal medium, whereas at initial time, 
this difference in the variations of magnetic field in two 
mediums is less illustrious.

As we can see from figures \ref{eb.iso1}, \ref{eb.iso2} and 
\ref{eb.aniso}, the decay of magnetic field with time is 
very slow in conducting medium and it nearly remains strong. 
So, it is plausible to explore the effect of strong magnetic 
field through an anisotropy, created by it.

\subsubsection{Strong magnetic field-induced anisotropy}
In the presence of a magnetic field, the quark momentum
$\mathbf{p}$ gets decomposed into the transverse and
longitudinal components with respect to the direction of magnetic field
(say, $3$-direction), hence the dispersion relation for the
quark of $i$-th flavor is modified quantum mechanically into
\begin{eqnarray}\label{dispersion relation}
\omega_{i,n}(p_L)=\sqrt{p_L^2+m_i^2+2n\left|q_iB\right|}
~,\end{eqnarray}
where $n=0$, $1$, $2$,$\cdots$ are the quantum numbers
to specify the Landau levels. But in the SMF limit ($eB \gg T^2$), the 
quarks are rarely excited thermally to the
higher Landau levels due to very high energy gap 
($\sim \mathcal{O}(\sqrt{eB})$, hence $p_T$ becomes much
smaller than $p_L$ which results in a momentum anisotropy. Thus
unlike the earlier one due to the asymptotic expansion,
the anisotropic parameter, $\xi$ becomes negative.

Like the earlier case, for weak anisotropy, the distribution function for 
quarks can be approximated from the isotropic one, except that here lower 
momentum particles are effectively removed from the distribution 
due to the negative value of $\xi$,
\be\label{A.D.F.(eB)}
f_{{\rm B},i}^{\rm aniso}(\mathbf{p^\prime};T)=\frac{1}{e^{\beta\sqrt{{{\rm p}^\prime}^2+\xi(\mathbf{p^\prime}\cdot\mathbf{n})^2+m_i^2}}+1}
~.\ee
Denoting the momentum vector in strong magnetic field limit ($p_T =0$) by 
$\mathbf{p^\prime}=(0,0,p_3)$ and assuming the direction of anisotropy 
along the direction of magnetic field, the above distribution function, for 
very small $\xi$, can be expanded as
\be
f_{{\rm B},i}^{\rm aniso}=f_i^{\xi=0}-\frac{\xi\beta p_3^2}{2\omega_i}f_i^{\xi=0}(1-f_i^{\xi=0})
~,\ee
where $\xi$-independent part of the quark distribution function in the 
presence of strong magnetic field is given by
\be
f_i^{\xi=0}=\frac{1}{e^{\beta\omega_i}+1}
~,\ee
where $\omega_i$ will be given from the dispersion relation 
\eqref{dispersion relation} in SMF limit ($n=0$) after 
identifying $p_L$ with $p_3$, {\em i.e.} $\omega_i =\sqrt{p_3^2+m_i^2}$.

In the SMF limit, the quark momentum is assumed to be purely longitudinal 
\cite{Gusynin:PLB450'1999,Rath:JHEP1712'2017,Bjorken expansion}. Therefore 
when the thermal medium is disturbed infinitesimally by an electric field, 
an electromagnetic current is induced in the longitudinal direction (3-direction) as
\begin{eqnarray}\label{current(eB1)}
J_3 = \sum_i q_i g_i \int\frac{d^3\rm{p}}{(2\pi)^3\omega_i}
p_3 [\delta f_i^q(\tilde{x},\tilde{p})+\delta f_i^{\bar q}(\tilde{x},\tilde{p})]
~,\end{eqnarray}
unlike $\mathbf{J}$ in the absence of magnetic field. 
In eq. (\ref{current(eB1)}), we have used new notations 
relevant to the calculations in SMF limit as 
$\tilde{x}=(x_0,0,0,x_3)$ and 
$\tilde{p}=(p_0,0,0,p_3)$. In this case, the 
electrical conductivity can be obtained from the third 
component of current in Ohm's law,
\be
J_3=\sigma_{\rm el} E_3
~.\ee
Due to dimensional reduction in the presence of a strong 
magnetic field, the density of states in two spatial directions 
perpendicular to the direction of magnetic field can be written 
in terms of $|q_iB|$ and as a result, the (integration) phase factor 
gets modified \cite{Gusynin:NPB462'1996,Bruckmann:PRD96'2017} as
\be\label{phase factor}
\int\frac{d^3{\rm p}}{(2\pi)^3}\rightarrow\frac{|q_iB|}{2\pi}\int \frac{dp_3}{2\pi}
~.\ee

The infinitesimal perturbation due to the 
action of external magnetic field is obtained from the relativistic 
Boltzmann transport equation in RTA, in conjunction with the strong 
magnetic field limit,
\be\label{R.B.T.E.(eB)}
p^0\frac{\partial f_i}{\partial x^0}+p^3\frac{\partial f_i}{\partial x^3}+q_i F^{03}p_3 \frac{\partial f_i}{\partial p^0}+q_i F^{30}p_0 \frac{\partial f_i}{\partial p^3}=-\frac{p_0}{\tau^B_i}\delta f_i
~,\ee
where $\tau^B_i$ denotes the relaxation time for quark in the 
presence of strong magnetic field. In the LLL approximation, the 
momentum-dependent relaxation-time is calculated \cite{Hattori:PRD95'2017} as
\be
\tau^B_i=\frac{\omega_i\left(e^{\beta\omega_i}-1\right)}{\alpha_sC_2m_i^2\left(e^{\beta\omega_i}+1\right)}\left[1\Bigg{/}\left\lbrace\int dp^\prime_3\frac{1}{\omega^\prime_i\left(e^{\beta\omega^\prime_i}+1\right)}\right\rbrace\right]
,\ee
where $C_2$ is the Casimir factor and the primed notations are 
used for antiquark. Now solving the RBTE (\ref{R.B.T.E.(eB)}) 
for the anisotropic distribution function, we obtain $\delta f_i$ as
\be
\nonumber\delta f_i &=& \frac{2\tau^B_i \beta q_iE_3p_3}{\omega_i}\left[f_i^{\xi=0}(1-f_i^{\xi=0})+\frac{\xi}{2}\left\lbrace -\frac{\beta p_3^2}{\omega_i}f_i^{\xi=0}(1-f_i^{\xi=0})+\frac{2\beta p_3^2}{\omega_i}{f_i^{\xi=0}}^2(1-f_i^{\xi=0})\right.\right. \\ && \left.\left.\hspace{4.9 cm}-\frac{p_3^2}{\omega_i^2}f_i^{\xi=0}(1-f_i^{\xi=0})+f_i^{\xi=0}(1-f_i^{\xi=0}) \right\rbrace\right]
.\ee
After substituting $\delta f_i$ in eq. (\ref{current(eB1)}), 
we get the electrical conductivity in the presence of a strong 
magnetic field-driven anisotropy,
\be\label{A.E.C.(eB)}
\nonumber\sigma_{\rm el,B}^{\rm aniso} &=& \frac{\beta}{\pi^2}\sum_i g_i q_i^2~|q_iB|\int dp_3~\frac{p_3^2}{\omega_i^2} ~ \tau_i^B ~ f_i^{\xi=0}(1-f_i^{\xi=0}) \\ && - \nonumber\frac{\xi\beta^2}{2\pi^2}\sum_i g_i q_i^2~|q_iB|\int d{p_3}~\frac{p_3^4}{\omega_i^3}~ \tau_i^B ~ f_i^{\xi=0}(1-f_i^{\xi=0})\left\lbrace 1-2f_i^{\xi=0}+\frac{1}{\beta\omega_i} \right\rbrace \\ && +\frac{\xi\beta}{2\pi^2}\sum_i g_i q_i^2~|q_iB|\int d{p_3}~\frac{p_3^2}{\omega_i^2}~ \tau_i^B ~ f_i^{\xi=0}(1-f_i^{\xi=0})
~,\ee
which can further be decomposed into 
\be\label{A.E.C.(1eB)}
\nonumber\sigma_{\rm el,B}^{\rm aniso} &=& \sigma_{\rm el}^{\xi=0}+\sigma_{\rm el}^{\xi\ne 0} \\ &=& \nonumber\sigma_{\rm el}^{\xi=0} - \xi\left[\frac{\beta^2}{2\pi^2}\sum_i g_i q_i^2~|q_iB|\int d{p_3}~\frac{p_3^4}{\omega_i^3}~ \tau_i^B ~ f_i^{\xi=0}(1-f_i^{\xi=0})\left\lbrace 1-2f_i^{\xi=0}+\frac{1}{\beta\omega_i} \right\rbrace\right. \\ && \left.-\frac{\beta}{2\pi^2}\sum_i g_i q_i^2~|q_iB|\int d{p_3}~\frac{p_3^2}{\omega_i^2}~ \tau_i^B ~ f_i^{\xi=0}(1-f_i^{\xi=0})\right]
.\ee

\begin{figure}[]
\begin{center}
\begin{tabular}{c c}
\includegraphics[width=7.4cm]{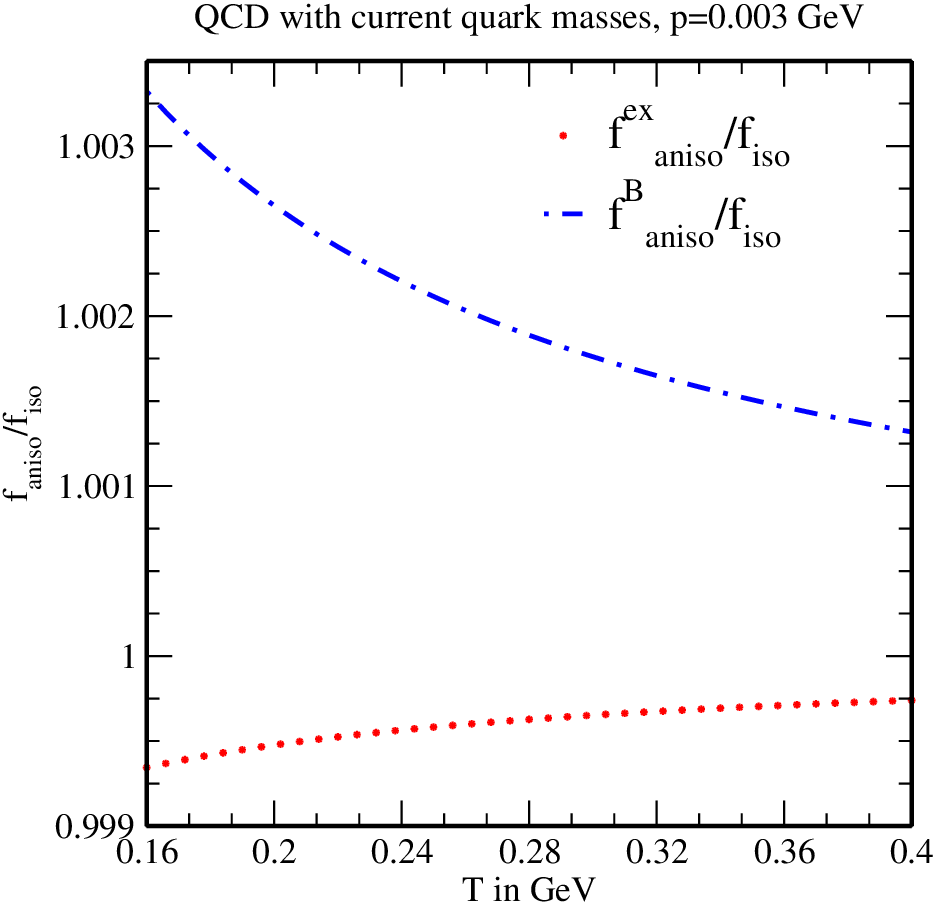}&
\hspace{0.6 cm}
\includegraphics[width=7.4cm]{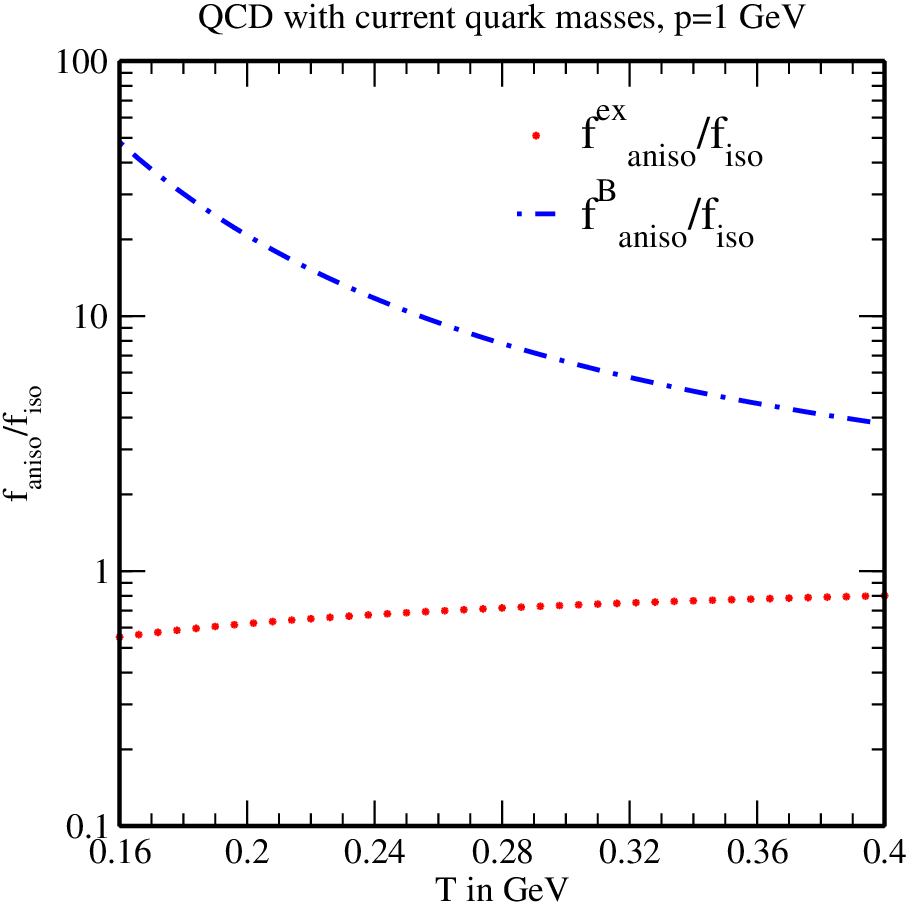} \\
a & b
\end{tabular}
\caption{Variation of the ratio $f_{\rm aniso}/f_{\rm iso}$ with temperature 
in the presence of momentum anisotropies both due to 
asymptotic expansion and strong magnetic field ($15$ $m_\pi^2$) at (a) low 
momentum and (b) high momentum, where the current quark mass has been used.}\label{fu.ideal}
\end{center}
\end{figure}

\begin{figure}[]
\begin{center}
\begin{tabular}{c c}
\includegraphics[width=7.3cm]{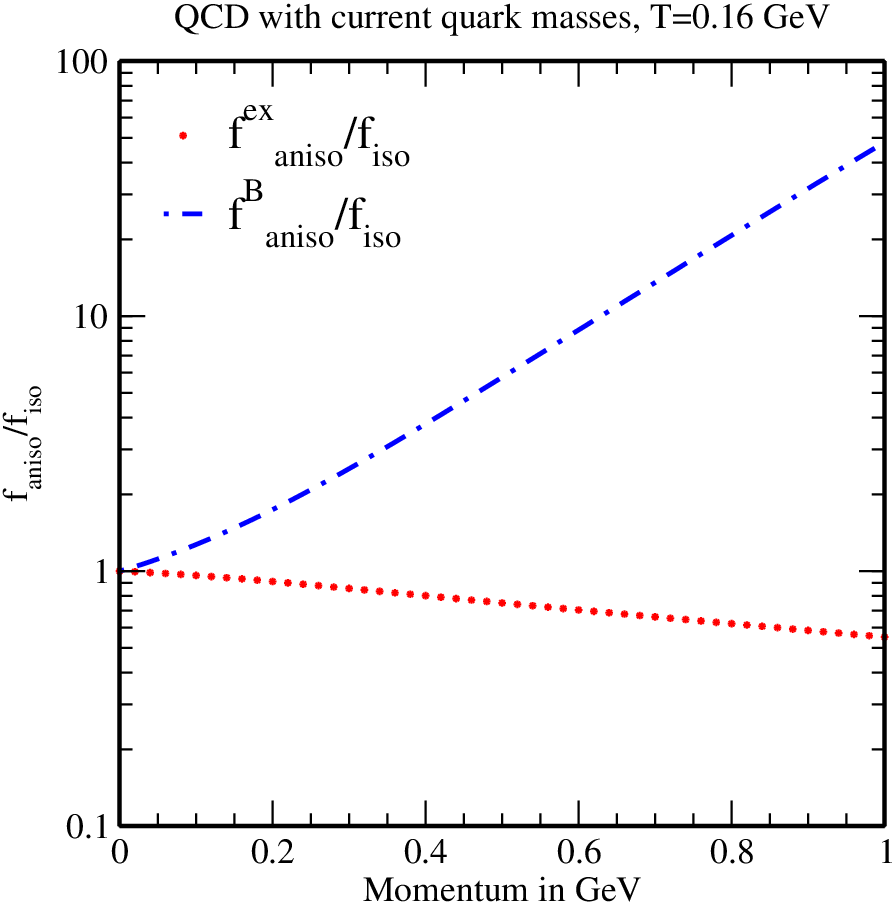}&
\hspace{0.67 cm}
\includegraphics[width=7.3cm]{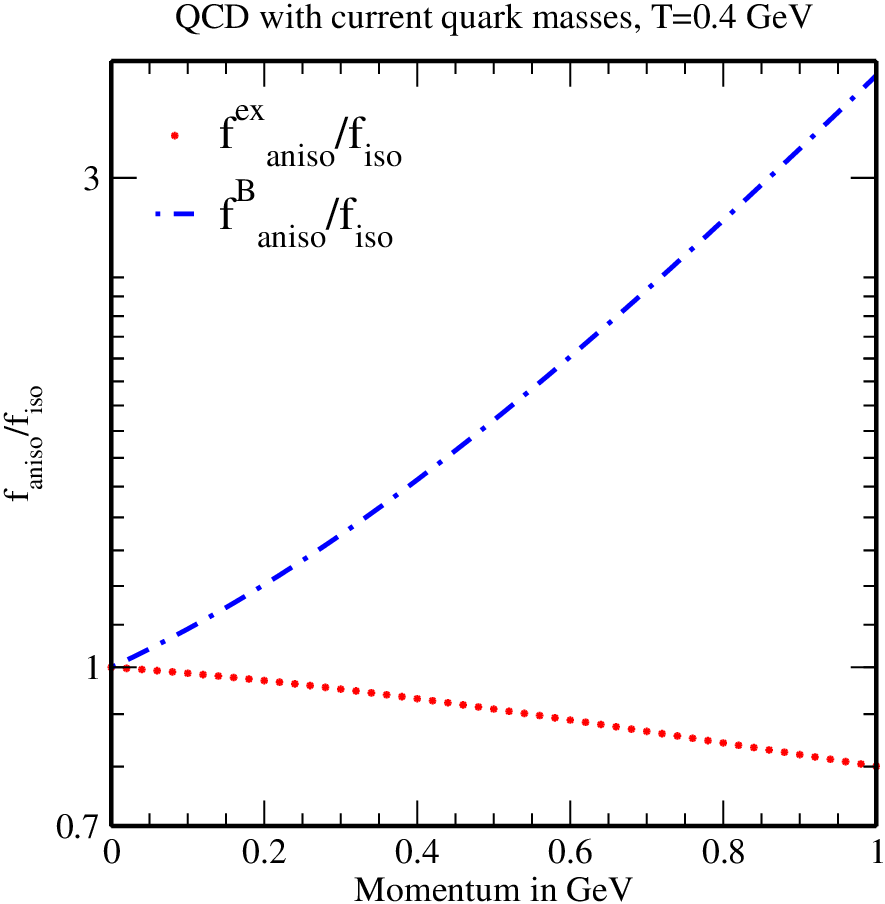} \\
a & b
\end{tabular}
\caption{Variation of the ratio $f_{\rm aniso}/f_{\rm iso}$ with momentum 
in the presence of momentum anisotropies both due to asymptotic 
expansion and strong magnetic field ($15$ $m_\pi^2$) at (a) low temperature 
and (b) high temperature, where the current quark mass has been used.}\label{fup.ideal}
\end{center}
\end{figure}

Before analysing the results on the electrical conductivity in 
the presence of anisotropies arising either due to the expansion or due to 
the strong magnetic field, we wish to understand first how the distribution 
function in an isotropic medium gets affected in the presence of anisotropies, 
because, in kinetic theory approach, the conductivities are mainly affected 
by the distribution function embodying the effects of anisotropy, the 
phase-space factor and the relaxation time. Therefore, we must understand how 
the ratios, $\dex/\df$, $\db/\df$ depend on the temperature at low and high 
momenta or {\em vice versa}, 
which are numerically plotted in figures \ref{fu.ideal} and \ref{fup.ideal}, 
respectively. The observations in the above figures can be readily understood 
by an order of estimate for the ratios for weak-momentum anisotropy 
($\xi < 1$) for nearly massless $u$ quark: $\dex/\df \sim e^{-c\frac{p}{T}}$, $\db/\df\sim e^{+c^\prime\frac{p}{T}}$, in both low and high momentum limits, with the  constant, $c<c^\prime <1$. The crucial negative and positive signs 
in exponentials arise due to the positive and negative 
anisotropic parameter, in expansion-driven and magnetic 
field-driven cases, respectively.

Let us start with the variation of $\dex/\df$ with $T$ in low momentum 
regime (figure \ref{fu.ideal}a). As $T$ increases, $p/T$ decreases, resulting 
an increase in $\dex/\df$ due to the lesser Boltzmann damping and 
an obvious decrease in $\db/\df$. The slower and relative faster variations 
are due to the smaller value of $c$ with respect to $c^\prime$. 
For higher momentum the variations (in figure \ref{fu.ideal}b) as well as the 
magnitudes of the ratios are more pronounced. The variations of the 
ratios with momentum at a fixed temperature (in figures \ref{fup.ideal}a 
and \ref{fup.ideal}b) are much easier to understand because the variable ($p$) in the  exponential is proportional to $p/T$, hence the variations become 
just opposite to the variation with temperature in figure \ref{fu.ideal}.

Before proceeding to discuss the results, it is to be mentioned that
we can not take arbitrarily large value of temperature 
due to the constraint of SMF limit ($eB \gg T^2$). For example, while 
computing the electrical conductivity as a function of temperature 
in a magnetic field-driven anisotropy with a strong magnetic field,
$eB=15$ $m_\pi^2$ ($m_\pi^2 \sim 0.02 ~ {\rm GeV}^2$), the temperature 
can be increased from $T_c$  up to 0.4 GeV. If the magnetic field
becomes even stronger, the temperature can go higher within the SMF limit.

\begin{figure}[]
\begin{center}
\includegraphics[width=7.9cm]{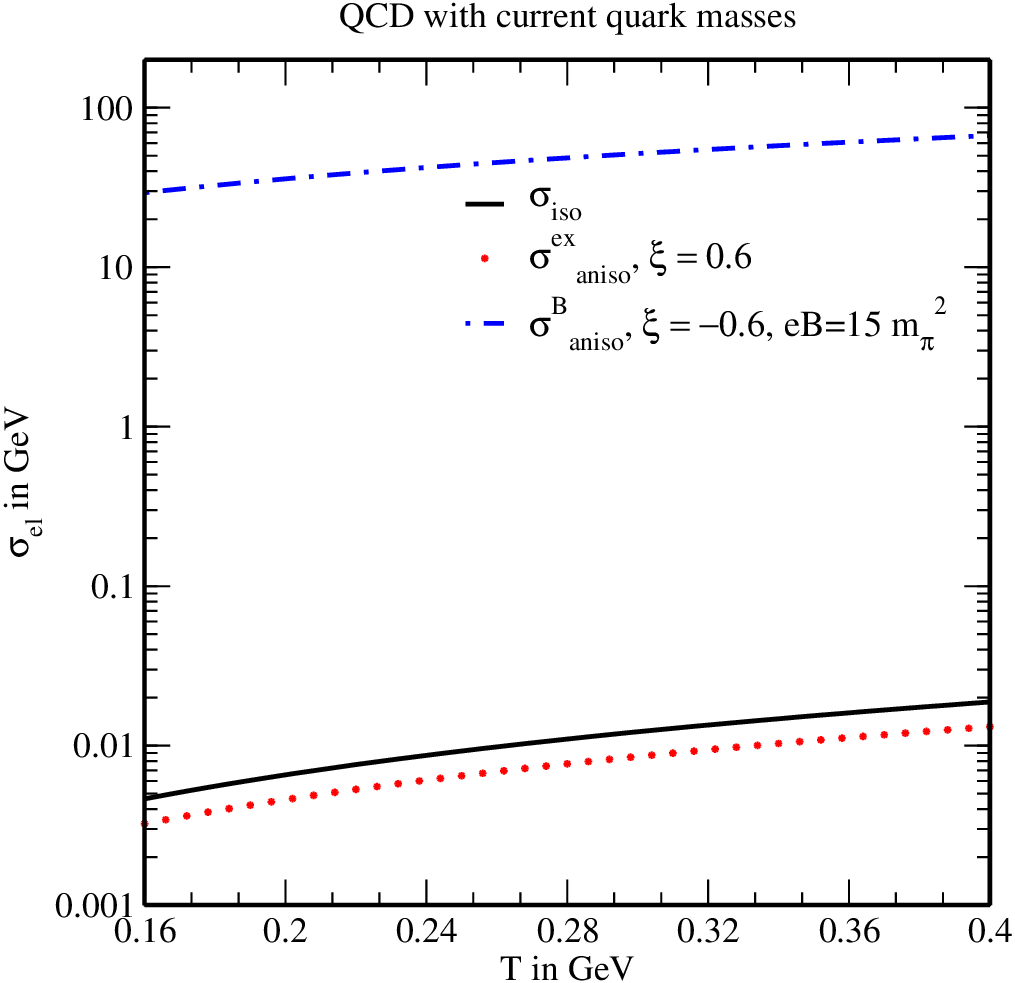}
\caption{Variation of electrical conductivity 
with temperature in the presence of momentum anisotropies 
both due to asymptotic expansion and strong 
magnetic field ($15$ $m_\pi^2$), where the current quark 
masses have been used.}\label{el.ideal}
\end{center}
\end{figure}

The above observations on the distribution functions facilitate to 
understand the results on the electrical conductivity for a thermal 
QCD medium with three flavors ($u$, $d$ and $s$) with their current masses 
in figure \ref{el.ideal}. For the isotropic medium (denoted by solid line), 
$\ec$ increases with the temperature, whereas due to the insertion 
of weak-momentum anisotropy (labelled as dotted line), $\ec$ gets slightly 
decreased because the ratio $\dex/\df$ is always less than 1 
for the entire range of temperature (as in figure \ref{fu.ideal}a). 
On the other hand, the relative magnitude of $\ec$ in magnetic 
field-driven anisotropic medium (labelled as dashed-dotted line) becomes very large
due to relatively large ratio, $\db/\df$. 
However $\ec$ increases with $T$, albeit  
the ratio, $\db/\df$ decreases with temperature 
(as in figure \ref{fu.ideal}). The decrease 
in $\db/\df$ at high temperature becomes much slower 
and approaches towards unity, but the phase-space factor ($\sim|q_iB|$)
and the relaxation time in SMF limit together 
compensate the minimal decrease in $\db$ and give an overall 
increasing trend in $\ec$ in the presence of strong magnetic field. The 
large value of $\ec$ in the strong magnetic field regime arises 
due to the large relaxation-time ($\tau^B$), because it is inversely 
proportional to the square of the mass, where the current quark mass is 
very small. Recently, similar results have been found in 
\cite{Hattori:PRD94'2016}, where $\sigma_{\rm el}$ is calculated in the 
diagrammatic method in the strong magnetic field regime and its large 
value is due to the smaller value of the current quark masses. This 
motivates us to recalculate the electrical conductivity with the 
quasiparticle masses in subsection 5.1.

\section{Thermal conductivity}
This section is devoted to the determination of the thermal conductivity 
of a hot QCD medium using the relativistic Boltzmann transport equation. 
In non-relativistic case, the heat equation is obtained by the validity 
of the first and second laws of thermodynamics, where the flow of heat 
is proportional to the temperature gradient and the proportionality 
factor is called the thermal conductivity. The heat does not flow 
directly, but it diffuses, depending on the internal structure of the 
medium it travels through. Similarly for a relativistic QCD system, the 
behavior of heat flow depends on the features of the medium. Thermal 
conductivity of a particular medium helps to describe the heat flow in that 
medium and it may leave significant effects on the hydrodynamic evolution of 
the systems with nonzero baryon chemical potential. To see how the heat flow 
gets affected, we have calculated thermal conductivity for both isotropic and 
anisotropic hot QCD mediums in subsections 3.1 and 3.2, respectively.

\subsection{Thermal conductivity for an isotropic thermal medium}
Heat flow four-vector is defined as the difference between the energy 
diffusion and the enthalpy diffusion,
\be\label{heat flow}
Q_\mu=\Delta_{\mu\alpha}T^{\alpha\beta}u_\beta-h\Delta_{\mu\alpha}N^\alpha
,\ee
where $\Delta_{\mu\alpha}=g_{\mu\alpha}-u_\mu u_\alpha$ is the projection 
operator, $h$ is the enthalpy per particle which in terms of energy density, 
pressure and particle number density is represented as 
$h=(\varepsilon+P)/n$, $T^{\alpha\beta}$ denotes the 
energy-momentum tensor, and $N^\alpha$ is the particle flow 
four-vector. $N^\alpha$ and $T^{\alpha\beta}$ are also known 
as the first and second moments of the distribution function, 
respectively, with the following expressions,
\be
&&N^\alpha=\sum_i 2 g_i\int \frac{d^3{\rm p}}{(2\pi)^3\omega_i}p^\alpha f_i(x,p) ~ \label{P.F.F.}, \\ &&T^{\alpha\beta}=\sum_i 2 g_i\int \frac{d^3{\rm p}}{(2\pi)^3\omega_i}p^\alpha p^\beta f_i(x,p) 
~ \label{E.M.T.}.\ee
It is also possible to obtain the particle number density from 
eq. (\ref{P.F.F.}), the energy density and the pressure from 
eq. (\ref{E.M.T.}) as 
$n=N^\alpha u_\alpha$, $\varepsilon=u_\alpha T^{\alpha\beta} u_\beta$ 
and $P=-\Delta_{\alpha\beta}T^{\alpha\beta}/3$, respectively. From 
equations (\ref{heat flow}), (\ref{P.F.F.}) and (\ref{E.M.T.}), one 
can find that in the rest frame of the heat bath or fluid, heat flow 
four-vector is orthogonal to the fluid four-velocity, {\em i.e.}
\be
Q_\mu u^\mu=0
~.\ee
Thus in the rest frame of the fluid, the heat flow is purely 
spatial and this component of heat flow due to the action of 
external disturbances can be written, in terms of the 
nonequilibrium part of the distribution function as
\be\label{heat1}
\mathbf{Q}=\sum_i 2 g_i\int \frac{d^3{\rm p}}{(2\pi)^3} ~ \frac{\mathbf{p}}{\omega_i}(\omega_i-h_i) \delta f_i(x,p)
~.\ee

In order to define the thermal conductivity for a system, the number 
of particles in that system must be conserved and therefore it 
requires the associated chemical potential to be nonzero. In the 
beginning of the universe and also in the initial stages of the 
heavy ion collisions, the value of chemical 
potential ($\mu$) is small but nonzero. In the Navier-Stokes equation, the 
heat flow is related to the thermal potential ($U=\mu/T$) \cite{Greif:PRE87'2013} as
\be\label{heat}
\nonumber Q_\mu &=& -\kappa\frac{nT^2}{\varepsilon+P}\nabla_\mu U \\ &=& \kappa\left[\nabla_\mu T - \frac{T}{\varepsilon+P}\nabla_\mu P\right] 
,\ee
where the coefficient $\kappa$ is known as the thermal 
conductivity and $\nabla_\mu=\partial_\mu-u_\mu u_\nu\partial^\nu$ is the 
four-gradient, which, in the rest frame of the heat bath, {\em i.e.} in the 
local rest frame, is replaced by $\partial_j$ (or $\partial/\partial x^j$). Thus, 
in the local rest frame, the spatial component of the heat flow is written as
\be\label{heat2}
\mathbf{Q}=-\kappa\left[\frac{\partial T}{\partial\mathbf{x}}-\frac{T}{nh}\frac{\partial P}{\partial\mathbf{x}}\right]
.\ee
The thermal conductivity ($\kappa$) can be determined by comparing 
equations (\ref{heat1}) and (\ref{heat2}), so we need to first 
find $\delta f_i$. In the local rest frame, the flow velocity and 
temperature depend on the spatial and temporal coordinates, so the 
distribution function can be expanded in terms of the gradients of 
flow velocity and temperature. Thus, the relativistic Boltzmann 
transport equation (\ref{R.B.T.E.(1)}) takes the following form,
\be\label{eq1}
p^\mu\partial_\mu T\frac{\partial f_i}{\partial T}+p^\mu\partial_\mu(p^\nu u_\nu)\frac{\partial f_i}{\partial p^0}+q_i\left[F^{0j}p_j\frac{\partial f_i}{\partial p^0}+F^{j0}p_0\frac{\partial f_i}{\partial p^j}\right]=-\frac{p^\nu u_\nu}{\tau_i}\delta f_i
~,\ee
where $f_i=f_i^{\rm iso}+\delta f_i$ and $p_0=\omega_i-\mu_i$, which, 
for very small value of $\mu_i$, can be approximated as 
$p_0\approx\omega_i$. After dropping out the infinitesimal 
correction to the local equilibrium distribution function 
($\delta f_i$) from the left hand side of eq. (\ref{eq1}) and then 
using the following partial derivatives,
\be
&&\frac{\partial f_i^{\rm iso}}{\partial T}=\frac{p_0}{T^2}f_i^{\rm iso}(1-f_i^{\rm iso}), \\ && \frac{\partial f_i^{\rm iso}}{\partial p^0}=-\frac{1}{T}f_i^{\rm iso}(1-f_i^{\rm iso}), \\ && \frac{\partial f_i^{\rm iso}}{\partial p^j}=-\frac{p^j}{Tp_0}f_i^{\rm iso}(1-f_i^{\rm iso})
,\ee
we solve eq. (\ref{eq1}) to get the infinitesimal disturbance,
\be\label{delta}
\nonumber\delta f_i &=& -\frac{\tau_i f_i^{\rm iso}(1-f_i^{\rm iso})}{p_0}\left[\frac{p_0}{T^2}\left\lbrace p^0\partial_0 T+p^j\partial_j T \right\rbrace-\frac{1}{T}\left\lbrace p^0\partial_0 p_0+p^j\partial_jp_0 \right\rbrace\right. \\ && \hspace{3 cm}\left.\nonumber-\frac{1}{T}\left\lbrace p^0p^\nu\partial_0 u_\nu+p^j p^\nu\partial_j u_\nu \right\rbrace-\frac{2q_i}{T}\mathbf{E}\cdot\mathbf{p}\right] \\ &=& \nonumber-\frac{\tau_i f_i^{\rm iso}(1-f_i^{\rm iso})}{T}\left[\frac{p_0}{T}\partial_0 T+\frac{1}{T}p^j\partial_j T+T\partial_0\left(\frac{\mu}{T}\right)+\frac{T}{p_0}p^j\partial_j\left(\frac{\mu}{T}\right)\right. \\ && \hspace{3.3 cm}\left.-p^\nu\partial_0 u_\nu -\frac{p^jp^\nu}{p_0}\partial_j u_\nu - \frac{2q_i}{p_0}\mathbf{E}\cdot\mathbf{p}\right]
.\ee
Substituting 
$\partial_j\left(\frac{\mu}{T}\right)=-\frac{h}{T^2}\left(\partial_jT-\frac{T}{nh}\partial_jP\right)$ and using 
$\partial_0 u_\nu=\nabla_\nu P/(nh)$ from the 
energy-momentum conservation, we get the 
final expression for $\delta f_i$ as
\be
\nonumber\delta f_i &=& -\frac{\tau_i f_i^{\rm iso}(1-f_i^{\rm iso})}{T}\left[\frac{p_0}{T}\partial_0 T+\left(\frac{p_0-h}{p_0}\right)\frac{p^j}{T}\left(\partial_jT-\frac{T}{nh}\partial_jP\right)+T\partial_0\left(\frac{\mu}{T}\right)\right. \\ && \hspace{3.3 cm}\left.-\frac{p^jp^\nu}{p_0}\partial_j u_\nu - \frac{2q_i}{p_0}\mathbf{E}\cdot\mathbf{p}\right]
.\ee
After substituting the $\delta f_i$ expression in 
eq. (\ref{heat1}) and then comparing it with 
eq. (\ref{heat2}), we get the thermal conductivity 
for the isotropic medium,
\be\label{iso.}
\kappa^{\rm iso} = \frac{\beta^2}{3\pi^2}\sum_ig_i\int d{\rm p} \frac{{\rm p}^4}{\omega_i^2}(\omega_i-h_i)^2 ~ \tau_i ~ f_i^{\rm iso}(1-f_i^{\rm iso})
~.\ee

\subsection{Thermal conductivity for an anisotropic thermal medium}
In this subsection we will first observe the effects due to the 
weak-momentum anisotropy on the thermal conductivity of hot QCD 
medium caused by the initial asymptotic expansion and then 
by the strong magnetic field as well.

\subsubsection{Expansion-induced anisotropy}
Using the Taylor series expansion of the anisotropic distribution 
function ($f^{\rm aniso}_{{\rm ex},i}$) up to the first order in $\xi$, the 
following partial derivatives have been calculated as
\be
&&\frac{\partial f_{{\rm ex},i}^{\rm aniso}}{\partial T}=\frac{p_0f_i^{\rm iso}(1-f_i^{\rm iso})}{T^2}-\frac{\xi (\mathbf{p}\cdot\mathbf{n})^2f_i^{\rm iso}(1-f_i^{\rm iso})}{2T^2p_0}\left[\frac{p_0}{T}-1-\frac{2p_0f_i^{\rm iso}}{T}\right], \\ && \frac{\partial f_{{\rm ex},i}^{\rm aniso}}{\partial p^0}=-\frac{f_i^{\rm iso}(1-f_i^{\rm iso})}{T}+\frac{\xi (\mathbf{p}\cdot\mathbf{n})^2f_i^{\rm iso}(1-f_i^{\rm iso})}{2Tp_0^2}\left[\frac{p_0}{T}+1-\frac{2p_0f_i^{\rm iso}}{T}\right], \\ && \nonumber\frac{\partial f_{{\rm ex},i}^{\rm aniso}}{\partial p^j}=-\frac{p^jf_i^{\rm iso}(1-f_i^{\rm iso})}{Tp_0}-\frac{\xi p^jc(\alpha,\theta,\phi)f_i^{\rm iso}(1-f_i^{\rm iso})}{2Tp_0} \\ && \hspace{7.9 cm}\times\left[2-\frac{{\rm p}^2}{p_0^2}-\frac{{\rm p}^2}{Tp_0}+\frac{2{\rm p}^2f_i^{\rm iso}}{Tp_0}\right]
,\ee
which are then substituted in eq. \eqref{eq1} to obtain $\delta f_i$,
\be
\nonumber\delta f_i &=& -\frac{\tau_i f_i^{\rm iso}(1-f_i^{\rm iso})}{T}\left[1-\frac{\xi(\mathbf{p}\cdot\mathbf{n})^2}{2p_0 T}+\frac{\xi(\mathbf{p}\cdot\mathbf{n})^2f_i^{\rm iso}}{p_0 T}\right] \\ && \nonumber\times\left[\frac{p_0}{T}\partial_0 T+\left(\frac{p_0-h_i}{p_0}\right)\frac{p^j}{T}\left(\partial_jT-\frac{T}{nh_i}\partial_jP\right)+T\partial_0\left(\frac{\mu}{T}\right)-\frac{p^jp^\nu}{p_0}\partial_j u_\nu\right] \\ && \nonumber-\frac{\tau_i f_i^{\rm iso}(1-f_i^{\rm iso})}{T}\frac{\xi(\mathbf{p}\cdot\mathbf{n})^2}{2p_0^2}\left[\frac{p_0}{T}\partial_0 T+\left(\frac{p_0+h_i}{p_0}\right)\frac{p^j}{T}\left(\partial_jT-\frac{T}{nh_i}\partial_jP\right)-T\partial_0\left(\frac{\mu}{T}\right)\right. \\ && \left.\nonumber+\frac{2p^j}{nh_i}\partial_jP+\frac{p^jp^\nu}{p_0}\partial_j u_\nu\right]+\frac{2q_i\tau_i}{p_0 T}\mathbf{E}\cdot\mathbf{p}f_i^{\rm iso}(1-f_i^{\rm iso}) \\ && \hspace{4.79 cm}\times\left[1+\frac{\xi(\mathbf{p}\cdot\mathbf{n})^2}{2p_0^2}\left\lbrace \frac{p_0^2}{\mathbf{p}^2}-1-\frac{p_0}{T}+\frac{2p_0f_i^{\rm iso}}{T} \right\rbrace\right]
.\ee
Now substituting the value of $\delta f_i$ in eq. (\ref{heat1}), 
we find the thermal conductivity for an expansion-driven 
anisotropic thermal QCD medium,
\be
\kappa_{\rm ex}^{\rm aniso} &=& \nonumber\frac{\beta^2}{3\pi^2}\sum_ig_i\int d{\rm p} ~ \frac{{\rm p}^4}{\omega_i^2}(\omega_i-h_i)^2 ~ \tau_i ~ f_i^{\rm iso}(1-f_i^{\rm iso}) \\ && \nonumber+\frac{\xi\beta^2}{18\pi^2}\sum_ig_i\int d{\rm p} ~ \frac{{\rm p}^6}{\omega_i^4}(\omega_i^2-h_i^2) ~ \tau_i ~ f_i^{\rm iso}(1-f_i^{\rm iso}) \\ && -\frac{\xi\beta^3}{18\pi^2}\sum_ig_i\int d{\rm p} ~ \frac{{\rm p}^6}{\omega_i^3}(\omega_i-h_i)^2 ~ \tau_i ~ f_i^{\rm iso}(1-2f_i^{\rm iso})(1-f_i^{\rm iso})
~,\ee
where the first expression in R.H.S. is the thermal conductivity 
for the isotropic thermal QCD medium. Thus one can write 
$\kappa_{\rm ex}^{\rm aniso}$ in terms of $\kappa^{\rm iso}$ as
\be\label{ex.}
\nonumber\kappa_{\rm ex}^{\rm aniso} &=& \kappa^{\rm iso}+\xi\left[\frac{\beta^2}{18\pi^2}\sum_ig_i\int d{\rm p} ~ \frac{{\rm p}^6}{\omega_i^4}(\omega_i^2-h_i^2) ~ \tau_i ~ f_i^{\rm iso}(1-f_i^{\rm iso})\right. \\ && \left.-\frac{\beta^3}{18\pi^2}\sum_ig_i\int d{\rm p} ~ \frac{{\rm p}^6}{\omega_i^3}(\omega_i-h_i)^2 ~ \tau_i ~ f_i^{\rm iso}(1-2f_i^{\rm iso})(1-f_i^{\rm iso})\right]
.\ee

We are now going to see how the thermal conductivity 
of the hot QCD medium gets modified due to the anisotropy 
developed by the strong magnetic field.

\subsubsection{Strong magnetic field-induced anisotropy}
The strong magnetic field restricts the dynamics of 
quarks to one spatial dimension, {\em i.e.} along the 
direction of magnetic field. So in the SMF limit, the 
spatial component of heat flow gets modified into
\be\label{heat(eb1)}
Q_3=\sum_i\frac{g_i|q_iB|}{2\pi^2}\int dp_3 ~ \frac{p_3}{\omega_i}(\omega_i-h_i^B) \delta f_i(\tilde{x},\tilde{p})
~.\ee
Similarly eq. (\ref{heat2}) takes the following form,
\be\label{heat(eb2)}
\nonumber Q_3 &=& -\kappa\left[\frac{\partial T}{\partial x_3}-\frac{T}{nh^B}\frac{\partial P}{\partial x_3}\right] \\ &=& \kappa\left[\partial_3T-\frac{T}{nh^B}\partial_3P\right]
,\ee
where $h^B=(\varepsilon+P)/n$ represents the 
enthalpy per particle in a strong magnetic field. For the 
charged particles in the SMF limit, the particle number 
density ($n$) is obtained from the following particle flow four-vector,
\be\label{P.F.F.(eb)}
N^\mu=\sum_i\frac{g_i|q_iB|}{2\pi^2}\int dp_3 \frac{\tilde{p}^\mu}{\omega_i} f_i(\tilde{x},\tilde{p}) 
~.\ee
The energy density ($\varepsilon$) and the 
pressure ($P$) are obtained from the following energy-momentum tensor,
\be\label{E.M.T.(eb)}
T^{\mu\nu}=\sum_i\frac{g_i|q_iB|}{2\pi^2}\int dp_3 \frac{\tilde{p}^\mu\tilde{p}^\nu}{\omega_i} f_i(\tilde{x},\tilde{p}) 
~.\ee
Now in terms of the gradients of flow velocity and temperature, the 
RBTE (\ref{R.B.T.E.(eB)}) in the presence of a strong magnetic 
field can be written as
\be\label{eq2}
\tilde{p}^\mu\frac{\partial T}{\partial \tilde{x}^\mu}\frac{\partial f_i}{\partial T}+\tilde{p}^\mu\frac{\partial (\tilde{p}^\nu u_\nu)}{\partial \tilde{x}^\mu}\frac{\partial f_i}{\partial p^0}+q_i\left[F^{03}p_3\frac{\partial f_i}{\partial p^0}+F^{30}p_0\frac{\partial f_i}{\partial p^3}\right]=-\frac{\tilde{p}^\nu u_\nu}{\tau_i^B}\delta f_i
~,\ee
where the variables, $\tilde{p}^\mu=(p^0,0,0,p^3)$ and 
$\tilde{x}^\mu=(x^0,0,0,x^3)$ are suited to the strong 
magnetic field calculation. Using the following partial derivatives,
\be
&&\frac{\partial f_{{\rm B},i}^{\rm aniso}}{\partial T}=\frac{p_0f_i^{\xi=0}(1-f_i^{\xi=0})}{T^2}-\frac{\xi p_3^2f_i^{\xi=0}(1-f_i^{\xi=0})}{2T^2p_0}\left[\frac{p_0}{T}-1-\frac{2p_0f_i^{\xi=0}}{T}\right], \\ && \frac{\partial f_{{\rm B},i}^{\rm aniso}}{\partial p^0}=-\frac{f_i^{\xi=0}(1-f_i^{\xi=0})}{T}+\frac{\xi p_3^2f_i^{\xi=0}(1-f_i^{\xi=0})}{2Tp_0^2}\left[\frac{p_0}{T}+1-\frac{2p_0f_i^{\xi=0}}{T}\right], \\ && \frac{\partial f_{{\rm B},i}^{\rm aniso}}{\partial p^3}=-\frac{p^3f_i^{\xi=0}(1-f_i^{\xi=0})}{Tp_0}-\frac{\xi p^3f_i^{\xi=0}(1-f_i^{\xi=0})}{2Tp_0} \\ && \hspace{7.9 cm}\times\left[2-\frac{p_3^2}{p_0^2}-\frac{p_3^2}{Tp_0}+\frac{2p_3^2f_i^{\xi=0}}{Tp_0}\right]
,\ee
we obtain $\delta f_i$ from eq. (\ref{eq2}),
\be
\nonumber\delta f_i &=& -\frac{\tau_i^B f_i^{\xi=0}(1-f_i^{\xi=0})}{T}\left[1-\frac{\xi p_3^2}{2p_0 T}+\frac{\xi p_3^2f_i^{\xi=0}}{p_0 T}\right] \\ && \nonumber\times\left[\frac{p_0}{T}\partial_0 T+\left(\frac{p_0-h_i^B}{p_0}\right)\frac{p^3}{T}\left(\partial_3T-\frac{T}{nh_i^B}\partial_3P\right)+T\partial_0\left(\frac{\mu}{T}\right)-\frac{p^3\tilde{p}^\nu}{p_0}\partial_3 u_\nu\right] \\ && \nonumber-\frac{\tau_i^B f_i^{\xi=0}(1-f_i^{\xi=0})}{T}\frac{\xi p_3^2}{2p_0^2}\left[\frac{p_0}{T}\partial_0 T+\left(\frac{p_0+h_i^B}{p_0}\right)\frac{p^3}{T}\left(\partial_3T-\frac{T}{nh_i^B}\partial_3P\right)-T\partial_0\left(\frac{\mu}{T}\right)\right. \\ && \left.\nonumber+\frac{2p^3}{nh_i^B}\partial_3P+\frac{p^3\tilde{p}^\nu}{p_0}\partial_3 u_\nu\right]+\frac{2q_i\tau_i^B}{p_0 T}E_3p_3f_i^{\xi=0}(1-f_i^{\xi=0}) \\ && \hspace{4.79 cm}\times\left[1+\frac{\xi p_3^2}{2p_0^2}\left\lbrace \frac{p_0^2}{p_3^2}-1-\frac{p_0}{T}+\frac{2p_0f_i^{\xi=0}}{T} \right\rbrace\right]
.\ee
After substituting $\delta f_i$ in eq. (\ref{heat(eb1)}), the thermal 
conductivity in a strong magnetic field-driven anisotropic medium is obtained,
\be
\kappa_{\rm B}^{\rm aniso} &=& \nonumber\frac{\beta^2}{2\pi^2}\sum_ig_i|q_iB|\int dp_3 ~ \frac{p_3^2}{\omega_i^2}(\omega_i-h_i^B)^2 ~ \tau_i^B ~ f_i^{\xi=0}(1-f_i^{\xi=0}) \\ && \nonumber+\frac{\xi\beta^2}{4\pi^2}\sum_ig_i|q_iB|\int dp_3 ~ \frac{p_3^4}{\omega_i^4}(\omega_i^2-{h_i^B}^2) ~ \tau_i^B ~ f_i^{\xi=0}(1-f_i^{\xi=0}) \\ && -\frac{\xi\beta^3}{4\pi^2}\sum_ig_i|q_iB|\int dp_3 ~ \frac{p_3^4}{\omega_i^3}(\omega_i-h_i^B)^2 ~ \tau_i^B ~ f_i^{\xi=0}(1-2f_i^{\xi=0})(1-f_i^{\xi=0})
~.\ee
Thus $\kappa_{\rm B}^{\rm aniso}$ can be rewritten in terms 
of $\xi$-independent and $\xi$-dependent parts as
\be\label{eb}
\nonumber\kappa_{\rm B}^{\rm aniso} &=& \kappa^{\xi=0}+\kappa^{\xi\ne 0} \\ &=& \nonumber\kappa^{\xi=0}+\xi\left[\frac{\beta^2}{4\pi^2}\sum_ig_i|q_iB|\int dp_3 ~ \frac{p_3^4}{\omega_i^4}(\omega_i^2-{h_i^B}^2) ~ \tau_i^B ~ f_i^{\xi=0}(1-f_i^{\xi=0})\right. \\ && \left.-\frac{\beta^3}{4\pi^2}\sum_ig_i|q_iB|\int dp_3 ~ \frac{p_3^4}{\omega_i^3}(\omega_i-h_i^B)^2 ~ \tau_i^B ~ f_i^{\xi=0}(1-2f_i^{\xi=0})(1-f_i^{\xi=0})\right]
.\ee

\begin{figure}[]
\begin{center}
\includegraphics[width=7.9cm]{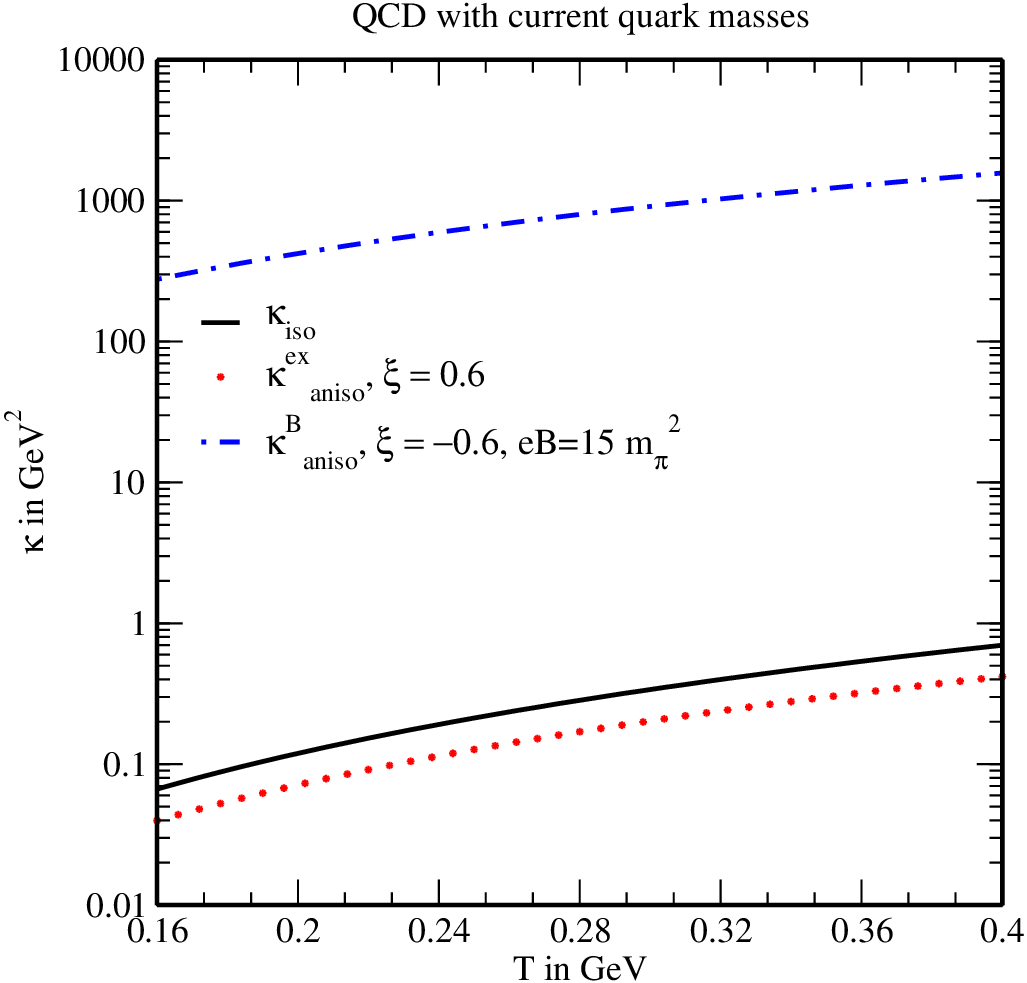}
\caption{Variation of thermal conductivity 
with temperature in the presence of momentum anisotropies 
both due to asymptotic expansion and strong 
magnetic field ($15$ $m_\pi^2$), where the current quark 
masses have been used.}\label{th.ideal}
\end{center}
\end{figure}

Figure \ref{th.ideal} depicts how the thermal conductivity 
varies with temperature for the isotropic medium and for the 
anisotropic mediums due to expansion-driven anisotropy and strong 
magnetic field-driven anisotropy. We have observed that 
$\kappa$ for the isotropic medium 
increases with the temperature. Similar increasing 
behavior of $\kappa$ is also noticed for the expansion-driven 
anisotropic medium, however its magnitude becomes smaller. 
If the origin of anisotropy is strong magnetic field, then the 
magnitude of $\kappa$ becomes unusually large. The above 
observations on the thermal conductivity could also be 
attributed to the behaviors of respective distribution functions, 
the phase-space factor and the relaxation time, where the 
last two factors are severely affected by the strong magnetic field 
only. This again necessitates the incorporation of the interactions among 
quarks through the quasiparticle model.

\section{Applications}
This section is devoted to study how the above behaviors observed 
in the electrical and thermal conductivities will help to understand
some specific properties of the medium. In subsection 4.1, we will 
observe how the interplay between the conductivities through the 
Wiedemann-Franz law gets modified in a thermal QCD medium in the 
presence of anisotropies arising due to different causes. In 
subsection 4.2, we will calculate the Knudsen number to have a say 
whether the thermal QCD medium is still in local equilibrium 
even in the presence of different anisotropies discussed hereinabove. 

\subsection{Wiedemann-Franz law}
\begin{figure}[]
\begin{center}
\includegraphics[width=7.9cm]{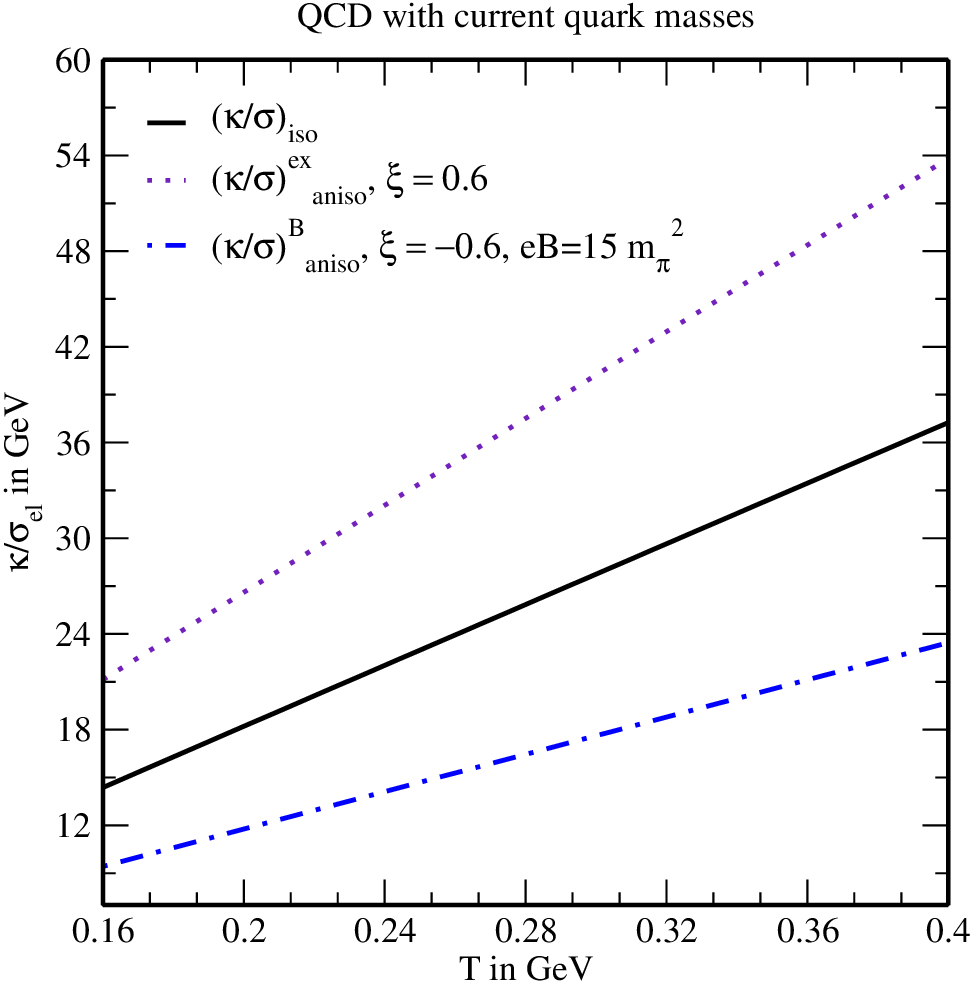}
\caption{Variation of the ratio of thermal conductivity to 
electrical conductivity with temperature in the presence 
of momentum anisotropies both due to asymptotic 
expansion and strong magnetic field ($15$ $m_\pi^2$), where the current quark 
masses have been used.}\label{wfl.ideal}
\end{center}
\end{figure}

According to the Wiedemann-Franz law, the ratio of charged 
particle contribution of the thermal conductivity to the electrical 
conductivity is proportional to the temperature,
\be
\frac{\kappa}{\sigma_{\rm el}}=LT
~,\ee
where the proportionality factor $L$ is called the Lorenz number. 
This law is perfectly satisfied by the matter which are good 
thermal and electrical conductors, such as metals. However for 
different cases, the deviation of the Wiedemann-Franz law has been 
observed, such as for the thermally populated electron-hole plasma 
in graphene, describing the signature of a Dirac fluid 
\cite{Crossno:S351'2016}, for the two-flavor quark matter in the 
Nambu-Jona-Lasinio (NJL) model \cite{Harutyunyan:PRD95'2017} and 
for the strongly interacting QGP medium \cite{Mitra:PRD96'2017}. 
In this work we intend to see how the Lorenz number for the 
thermal QCD matter varies by observing the ratio ($\kappa/\ec$) as a 
function of temperature in the presence of expansion-driven and strong 
magnetic-field driven anisotropies in figure \ref{wfl.ideal}.

In the isotropic medium, the ratio is found to 
increase linearly with temperature. When the isotropic 
medium is subjected to an expansion-driven anisotropy, 
$\kappa/\ec$ shows almost the same increasing behavior with 
temperature like in isotropic case, but its magnitude and 
slope ({\em i.e.} the Lorenz number) get enhanced. If the 
origin of anisotropy is strong magnetic field, then both the 
magnitude and the slope become smaller than the former 
descriptions. Thus in two different types of anisotropies we have 
observed nearly opposite behavior of $\kappa/\ec$, which 
can also be understood from the opposite behavior in 
electrical and thermal conductivities for the two 
aforesaid anisotropic mediums. This observation thus 
implies different Lorenz numbers ($\kappa/(\ec T)$) 
at the same temperature, depending on the anisotropies.

\subsection{Knudsen number}
The Knudsen number ($\Omega$) is required to be small 
for small deviation from equilibrium in the 
hydrodynamic regime, which is defined as
\be
\Omega=\frac{\lambda}{L}
~,\ee
where $\lambda$ denotes the mean free path and $L$ is the 
characteristic length scale of the system. One can calculate the mean 
free path by using the thermal conductivity ($\kappa$) of the medium,
\be
\lambda=\frac{3\kappa}{vC_V}
~,\ee
where $v$ is the relative speed and $C_V$ is the 
specific heat. Therefore the Knudsen number 
can be recast in terms of the thermal conductivity as
\be
\Omega=\frac{3\kappa}{LvC_V}
~.\ee
In the calculation we have taken $v\simeq 1$, $L=3$ fm, and $C_V$ is evaluated 
from the energy-momentum tensor, 
{\em i.e.} $C_V=\partial (u_\mu T^{\mu\nu} u_\nu)/\partial T$.

\begin{figure}[]
\begin{center}
\includegraphics[width=7.9cm]{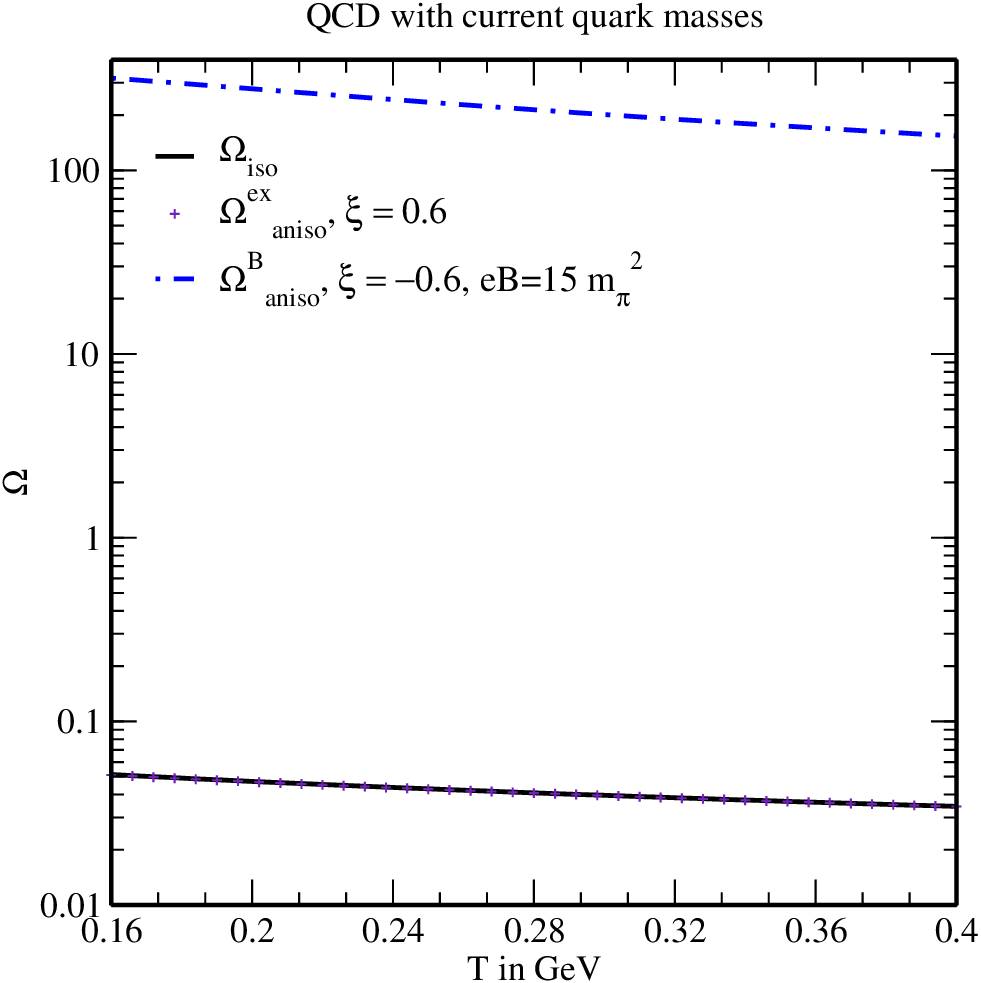}
\caption{Variation of the Knudsen number 
with temperature in the presence of momentum anisotropies 
both due to asymptotic expansion and strong 
magnetic field ($15$ $m_\pi^2$), where the current quark 
masses have been used.}\label{frac.ideal}
\end{center}
\end{figure}

In an isotropic medium, the Knudsen number decreases with the 
increase of temperature, which explains that the mean free path 
becomes much smaller than the characteristic length scale of the 
system. As a result, the medium approaches equilibrium 
faster. When the medium exhibits a weak-momentum anisotropy
due to the asymptotic expansion initially, the Knudsen number 
does not deviate considerably from its value in the isotropic 
medium (seen in figure \ref{frac.ideal}). However if the origin 
of anisotropy is the strong magnetic field ($eB=15$ $m_\pi^2$), a 
significant deviation from the isotropic one can be seen, where the Knudsen number 
has a larger magnitude (denoted as dashed-dotted line in figure \ref{frac.ideal}), 
which defies physical interpretation and urges us to use the 
quasiparticle model (seen in figure \ref{frac.aniso}).

\section{Quasiparticle description of hot QCD matter}
Till now, we, in fact, have not incorporated any interactions
among quarks and gluons in a thermal QCD medium either in the 
presence or absence of strong magnetic field. As a matter of 
fact, the magnitude and the variation of the electrical 
conductivity, thermal conductivity and Knudsen number become 
unrealistic. Hence we must resort to the quasiparticle 
description of particles, known as QPM, where different 
flavors acquire the medium generated masses, in addition to their 
current masses. The thermal mass is generated due to the 
interaction of quark with other particles of the medium, thus the 
quasiparticle model properly describes the collective properties 
of the medium. Earlier, this model was explained in different 
approaches such as the Nambu-Jona-Lasinio and PNJL based quasiparticle models \cite{Fukushima:PLB591'2004,Ghosh:PRD73'2006,Abuki:PLB676'2009}, quasiparticle 
model based on Gribov-Zwanziger quantization \cite{Su:PRL114'2015,Florkowski:PRC94'2016} 
etc. However, for our calculation, the effective mass (squared) of $i$-th 
flavor in a pure thermal medium is taken from \cite{Bannur:JHEP0709'2007},
\be\label{Q.P.M.}
m_i^2=m_{i0}^2+\sqrt{2}m_{i0}m_{iT}+m_{iT}^2
~,\ee
where $m_{i0}$ and $m_{iT}$ are the current quark mass and the thermally
generated mass of $i$-th flavor, respectively. The thermal mass
is calculated in one loop in refs. \cite{Braaten:PRD45'1992,Peshier:PRD66'2002} as
\be
m_{iT}^2=\frac{g^{\prime2}T^2}{6}
~,\ee
where $g^\prime$ is the running coupling that runs with the temperature of the 
medium. However, for a thermal medium in the presence of a strong 
magnetic field, the effective mass in thermal medium in eq. \eqref{Q.P.M.} can 
be generalized into
\be
\label{Q.P.M.(eb)}
m_i^2=m_{i0}^2+\sqrt{2}m_{i0}m_{iT,B}+m_{iT,B}^2
~.\ee
Like the evaluation of $m_{iT}$, $m_{iT,B}$ could be similarly derived 
from the self-consistent Schwinger-Dyson equation by the quark self-energy 
for a thermal QCD medium in a strong magnetic field, which needs to be 
evaluated now.

As we know that the quark self-energy is given by
\begin{eqnarray}\label{Q.S.E.}
\Sigma(p)=-\frac{4}{3} g^{2}i\int{\frac{d^4k}{(2\pi)^4}}\left[\gamma_\mu {S(k)}\gamma^\mu{D(p-k)}\right]
,\end{eqnarray}
where $4/3$ is the Casimir factor and $g$ is the running coupling that 
runs mainly with the magnetic field 
\cite{Ferrer:PRD91'2015,Andreichikov:PRL110'2013} because the magnetic 
field is the largest energy scale for quarks in the strong magnetic 
field regime. The quark propagator, $S(K)$ in vacuum is modified in the 
presence of magnetic field and is given \cite{Schwinger:PR82'1951,Tsai:PRD10'1974} 
by the Schwinger proper-time method in the momentum space,
\be\label{q. propagator}
S(k)=ie^{-\frac{k^2_\perp}{|q_iB|}}\frac{\left(\gamma^0 k_0-\gamma^3 k_z+m_i\right)}{k^2_\parallel-m^2_i}\left(1
-\gamma^0\gamma^3\gamma^5\right)
,\ee
where the four vectors are defined below with the metric tensors: 
$g^{\mu\nu}_\perp={\rm{diag}}(0,-1,-1,0)$ and $g^{\mu\nu}_\parallel= 
{\rm{diag}}(1,0,0,-1)$, 
\begin{eqnarray*}
&& k_{\perp\mu}\equiv(0,k_x,k_y,0), ~~ k_{\parallel\mu}\equiv(k_0,0,0,k_z)
~.\end{eqnarray*}
The gluon propagator in vacuum retains the same form even in the
presence of magnetic field, {\em i.e.}
\be
\label{g. propagator}
D^{\mu \nu} (p-k)=\frac{ig^{\mu \nu}}{(p-k)^2}
~.\ee

Next we obtain the form of quark and gluon propagators at finite temperature 
in the imaginary-time formalism and subsequently replace the energy integral 
($\int\frac{dp_0}{2\pi}$) by sums over Matsubara frequencies, to get the 
form of self-energy \eqref{Q.S.E.} at finite temperature. However, in a strong 
magnetic field along $z$-direction, the transverse component of the momentum 
becomes vanishingly small ($k_\perp \approx0$), so the exponential factor 
in eq. \eqref{q. propagator} becomes unity and the integration over the 
transverse component of the momentum becomes $|q_iB|$. Therefore the 
self-energy \eqref{Q.S.E.} at finite temperature in the SMF limit gets 
simplified into
\begin{eqnarray}\label{Q.S.E.(1)}
\nonumber\Sigma(p_\parallel) &=& \frac{2g^2}{3\pi^2}|q_iB|T\sum_n\int dk_z\frac{\left[\left(1+\gamma^0\gamma^3\gamma^5\right)\left(\gamma^0k_0
-\gamma^3k_z\right)-2m_i\right]}{\left[k_0^2-\omega^2_k\right]\left[(p_0-k_0)^2-\omega_{pk}^2\right]} \\ &=& \frac{2g^2|q_iB|}{3\pi^2}\int dk_z\left[(\gamma^0+\gamma^3\gamma^5)L^1-(\gamma^3+\gamma^0\gamma^5)k_zL^2\right]
,\end{eqnarray}
where $\omega^2_k=k_z^2+m_i^2$, $\omega_{pk}^2=(p_z-k_z)^2$ and 
$L^1$ and $L^2$ are the two frequency sums, which are given by
\be
&&L^1=T\sum_nk_0~\frac{1}{\left[k_0^2-\omega_k^2\right]}\frac{1}{\left[(p_0-k_0)^2-\omega_{pk}^2\right]} ~, \\ &&L^2=T\sum_n\frac{1}{\left[k_0^2-\omega_k^2\right]}\frac{1}{\left[(p_0-k_0)^2-\omega_{pk}^2\right]}
~.\ee
After calculating the above frequency sums \cite{Kapusta:BOOK'2006, 
Bellac:BOOK'1996}, the self-energy (\ref{Q.S.E.(1)}) can be simplified 
further into
\begin{equation}\label{Q.S.E.(2)}
\Sigma(p_\parallel)=\frac{g^2|q_iB|}{3\pi^2}\int \frac{dk_z}{\omega_k}\left[\frac{1}{e^{\beta\omega_k}-1}+\frac{1}{e^{\beta\omega_k}+1}\right]\left[\frac{\gamma^0p_0+\gamma^3p_z}{p_\parallel^2}+\frac{\gamma^0\gamma^5p_z+\gamma^3\gamma^5p_0}{p_\parallel^2}\right]
,\end{equation}
which yields the following form, after the integration over $k_z$,
\begin{eqnarray}\label{Q.S.E.(3)}
\Sigma(p_\parallel)=\frac{g^2|q_iB|}{3\pi^2}\left[\frac{\pi T}{2m_i}-\ln(2)\right]\left[\frac{\gamma^0p_0}{p_\parallel^2}+\frac{\gamma^3p_z}{p_\parallel^2}+\frac{\gamma^0\gamma^5p_z}{p_\parallel^2}+\frac{\gamma^3\gamma^5p_0}{p_\parallel^2}\right]
.\end{eqnarray}

Let us explore the covariant structure of the quark self-energy at 
finite temperature in an additional presence of magnetic field, which 
can in general be written \cite{Ayala:PRD91'2015,Karmakar:1902.02607} as
\begin{equation}\label{general q.s.e.}
\Sigma(p_\parallel)=A\gamma^\mu u_\mu+B\gamma^\mu b_\mu+C\gamma^5\gamma^\mu u_\mu+D\gamma^5\gamma^\mu b_\mu
~,\end{equation}
where $A$, $B$, $C$ and $D$ are the form factors, and $u^\mu$  (1,0,0,0) 
and $b^\mu$  (0,0,0,-1) specify the preferred directions of the heat bath 
and the magnetic field, respectively. These
vectors are responsible for breaking the Lorentz and rotational 
symmetries, respectively. We have obtained the form factors in LLL 
approximation as
\begin{eqnarray}
&&A=\frac{1}{4}{\rm Tr}\left[\Sigma\gamma^\mu u_\mu\right]=\frac{g^2|q_iB|}{3\pi^2}\left[\frac{\pi T}{2m_i}-\ln(2)\right]\frac{p_0}{p_\parallel^2} ~, \\ 
&&B=-\frac{1}{4}{\rm Tr}\left[\Sigma\gamma^\mu b_\mu\right]=\frac{g^2|q_iB|}{3\pi^2}\left[\frac{\pi T}{2m_i}-\ln(2)\right]\frac{p_z}{p_\parallel^2} ~, \\ 
&&C=\frac{1}{4}{\rm Tr}\left[\gamma^5\Sigma\gamma^\mu u_\mu\right]=-\frac{g^2|q_iB|}{3\pi^2}\left[\frac{\pi T}{2m_i}-\ln(2)\right]\frac{p_z}{p_\parallel^2} ~, \\ 
&&D=-\frac{1}{4}{\rm Tr}\left[\gamma^5\Sigma\gamma^\mu b_\mu\right]=-\frac{g^2|q_iB|}{3\pi^2}\left[\frac{\pi T}{2m_i}-\ln(2)\right]\frac{p_0}{p_\parallel^2}
~,\end{eqnarray}
which indicates that $C=-B$ and $D=-A$. 

The quark self-energy \eqref{general q.s.e.} can also 
be written in terms of the right-handed ($P_R=(1+\gamma^5)/2$) and 
left-handed ($P_L=(1-\gamma^5)/2$) chiral projection operators,
\begin{equation}\label{projection}
\Sigma(p_\parallel)=P_R\left[(A+C)\gamma^\mu u_\mu+(B+D)\gamma^\mu b_\mu
\right]P_L+P_L\left[(A-C)\gamma^\mu u_\mu+(B-D)\gamma^\mu b_\mu\right]P_R
~,\end{equation}
which becomes simplified as
\begin{equation}\label{projection1}
\Sigma(p_\parallel)=P_R\left[(A-B)\gamma^\mu u_\mu+(B-A)\gamma^\mu b_\mu
\right]P_L+P_L\left[(A+B)\gamma^\mu u_\mu+(B+A)\gamma^\mu b_\mu\right]P_R
~,\end{equation}
after the substitutions $C=-B$ and $D=-A$.

Therefore the effective quark propagator is obtained from the self-consistent 
Schwinger-Dyson equation in the presence of a strong magnetic field,
\be
S^{-1}(p_\parallel)=\gamma^\mu p_{\parallel\mu}-\Sigma(p_\parallel)
~,\ee
which in turn takes the following form in terms of projection operators,
\be
S^{-1}(p_\parallel)=P_R\gamma^\mu X_\mu P_L+P_L\gamma^\mu Y_\mu P_R
~,\ee
where
\begin{eqnarray}
&&\gamma^\mu X_\mu=\gamma^\mu p_{\parallel\mu}-(A-B)\gamma^\mu u_\mu-(B-A)\gamma^\mu b_\mu ~, \\ 
&&\gamma^\mu Y_\mu=\gamma^\mu p_{\parallel\mu}-(A+B)\gamma^\mu u_\mu-(B+A)\gamma^\mu b_\mu
~.\end{eqnarray}
Now the effective propagator is written as
\be
S(p_\parallel)=\frac{1}{2}\left[P_R\frac{\gamma^\mu Y_\mu}{Y^2/2}P_L+
P_L\frac{\gamma^\mu X_\mu}{X^2/2}P_R\right]
,\ee
where
\begin{eqnarray}
&&\frac{X^2}{2}=X_1^2=\frac{1}{2}\left[p_0-(A-B)\right]^2-\frac{1}{2}\left[p_z+(B-A)\right]^2 ~, \\ 
&&\frac{Y^2}{2}=Y_1^2=\frac{1}{2}\left[p_0-(A+B)\right]^2-\frac{1}{2}\left[p_z+(B+A)\right]^2
~.\end{eqnarray}

Thus the thermal mass (squared) at finite temperature and strong 
magnetic field is finally obtained by taking the $p_0=0, p_z\rightarrow 0$ 
limit of either $X_1^2$ or $Y_1^2$ (because both of them are equal),
\begin{eqnarray}\label{Mass}
m_{iT,B}^2=X_1^2\Big{|}_{p_0=0,p_z\rightarrow 0}=Y_1^2\Big{|}_{p_0=0,p_z\rightarrow 0}=\frac{g^2|q_iB|}{3\pi^2}\left[\frac{\pi T}{2m_i}-\ln(2)\right]
,\end{eqnarray}
which depends on both temperature and magnetic field.

\begin{figure}[]
\begin{center}
\begin{tabular}{c c}
\includegraphics[width=7.4cm]{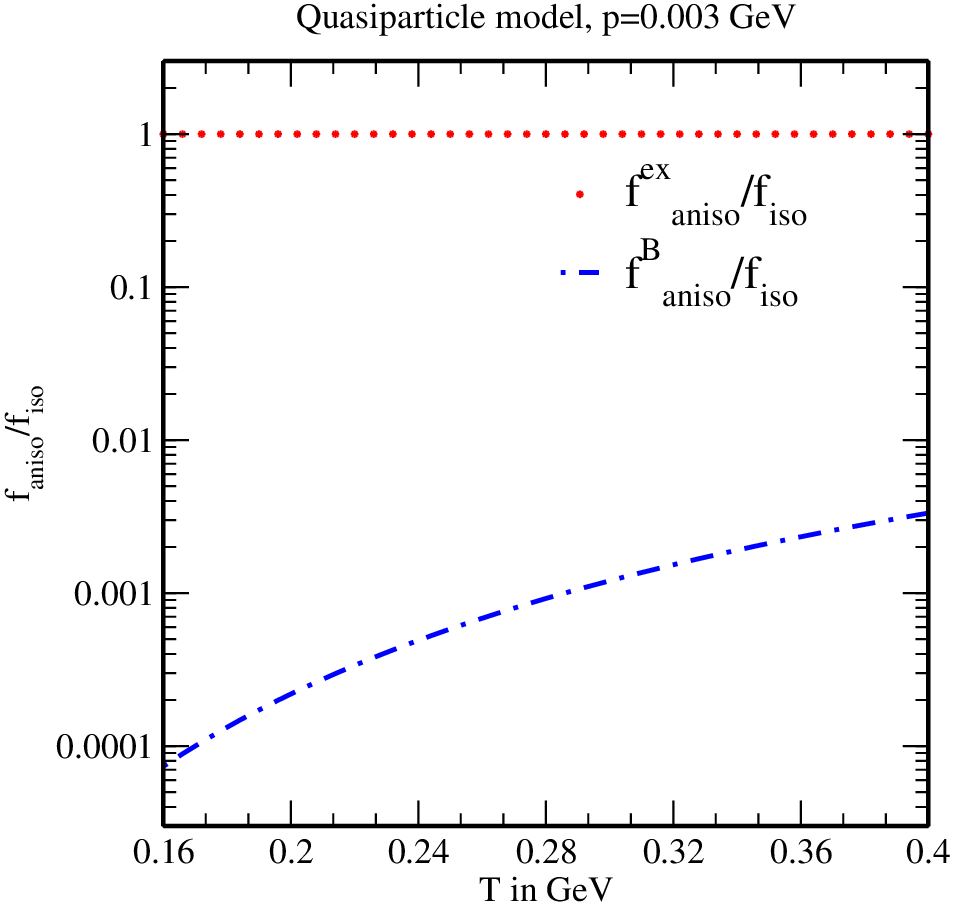}&
\hspace{0.6 cm}
\includegraphics[width=7.4cm]{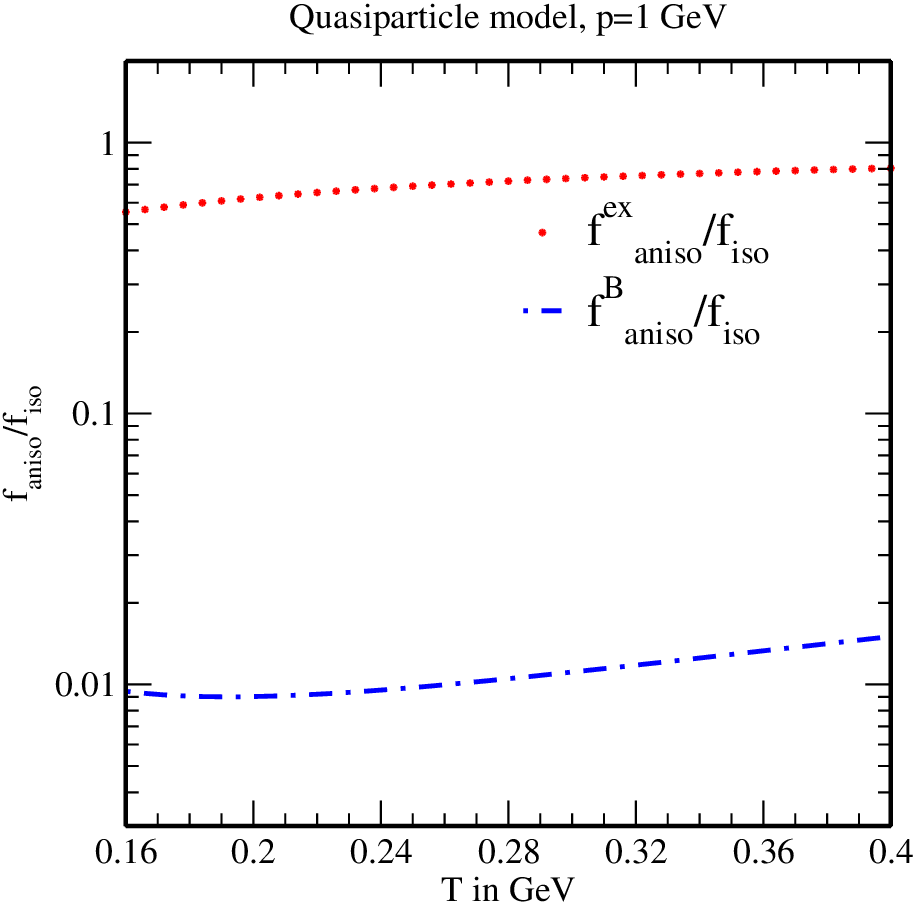} \\
a & b
\end{tabular}
\caption{Variation of the ratio $f_{\rm aniso}/f_{\rm iso}$ with temperature 
in the presence of momentum anisotropies both due to 
asymptotic expansion and strong magnetic field ($15$ $m_\pi^2$) at (a) low 
momentum and (b) high momentum, where the effective quark mass has been used.}\label{fu.qpm}
\end{center}
\end{figure}

\begin{figure}[]
\begin{center}
\begin{tabular}{c c}
\includegraphics[width=7.3cm]{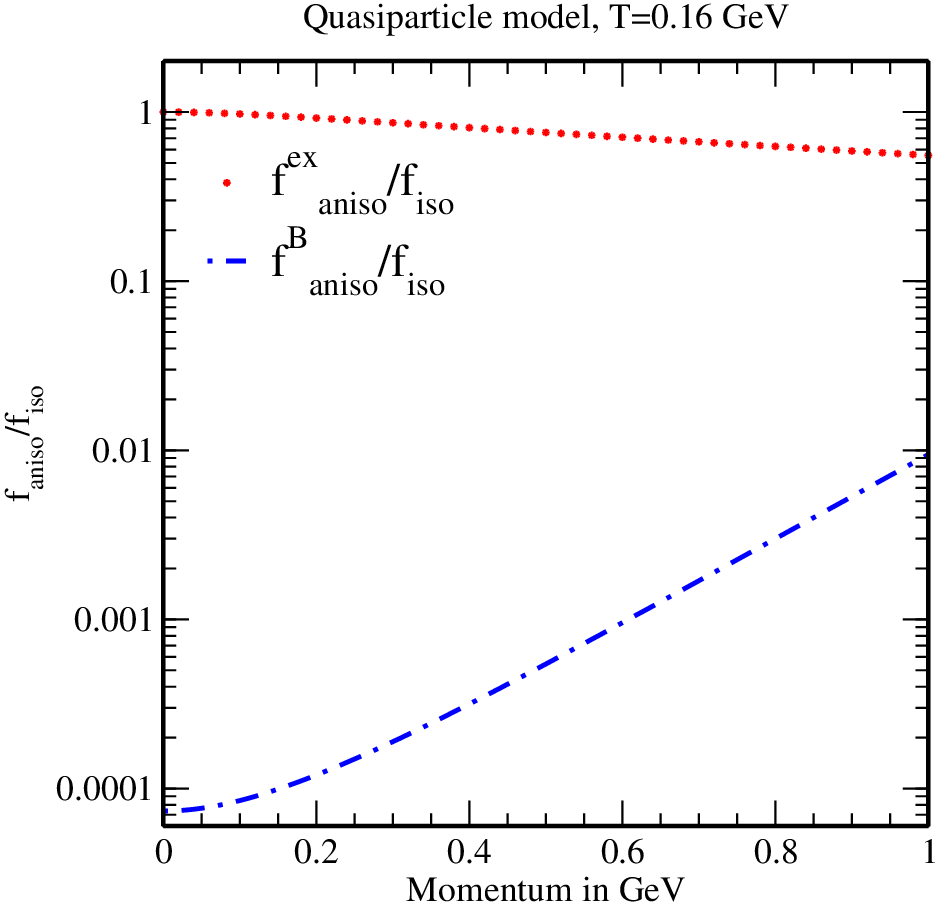}&
\hspace{0.67 cm}
\includegraphics[width=7.3cm]{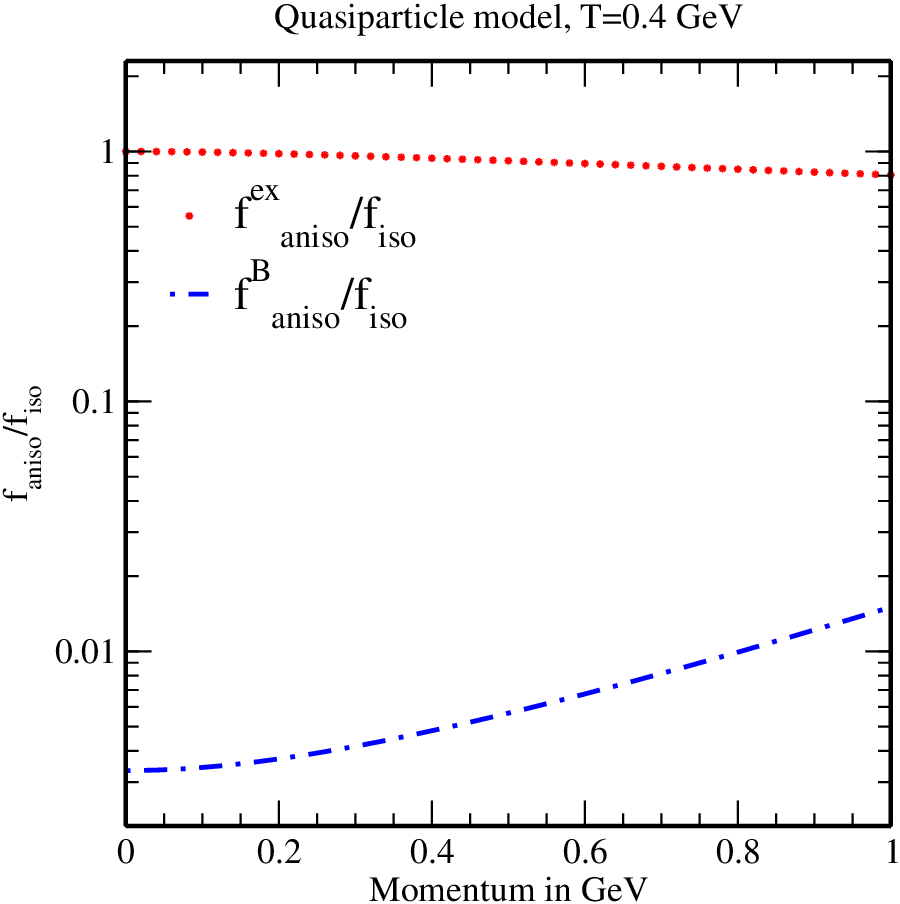} \\
a & b
\end{tabular}
\caption{Variation of the ratio $f_{\rm aniso}/f_{\rm iso}$ with momentum 
in the presence of momentum anisotropies both due to asymptotic 
expansion and strong magnetic field ($15$ $m_\pi^2$) at (a) low temperature 
and (b) high temperature, where the effective quark mass has been used.}\label{fup.qpm}
\end{center}
\end{figure}

In the quasiparticle description of particles, the distribution functions 
now contain the effective masses of the particles. Therefore, 
the distribution functions in the isotropic medium as well as in the expansion-driven 
anisotropic medium use the $T$-dependent effective mass \eqref{Q.P.M.}, 
whereas the  distribution function in the strong magnetic field-driven anisotropic 
medium uses the $T$ and $B$-dependent effective mass 
\eqref{Q.P.M.(eb)}. So, from figures \ref{fu.qpm} and 
\ref{fup.qpm}, we noticed that the behaviors of ratios 
($\dex/\df$ and $\db/\df$) get flipped in comparison to their respective 
behavior in ideal case 
(as in figures \ref{fu.ideal} and \ref{fup.ideal}). As the transport 
coefficients such as the electrical conductivity and the thermal conductivity are 
expressed in terms of the distribution function at finite temperature 
and/or magnetic field, so the knowledge about the behavior of distribution 
function in the QPM description is useful in understanding the transport 
properties of the hot QCD medium.

In the coming subsections we are going to discuss the results on the 
electrical conductivity, thermal conductivity and their applications 
using the quasiparticle model with three flavors ($u$, $d$ and $s$).

\subsection{Electrical conductivity}
\begin{figure}[]
\begin{center}
\includegraphics[width=7.9cm]{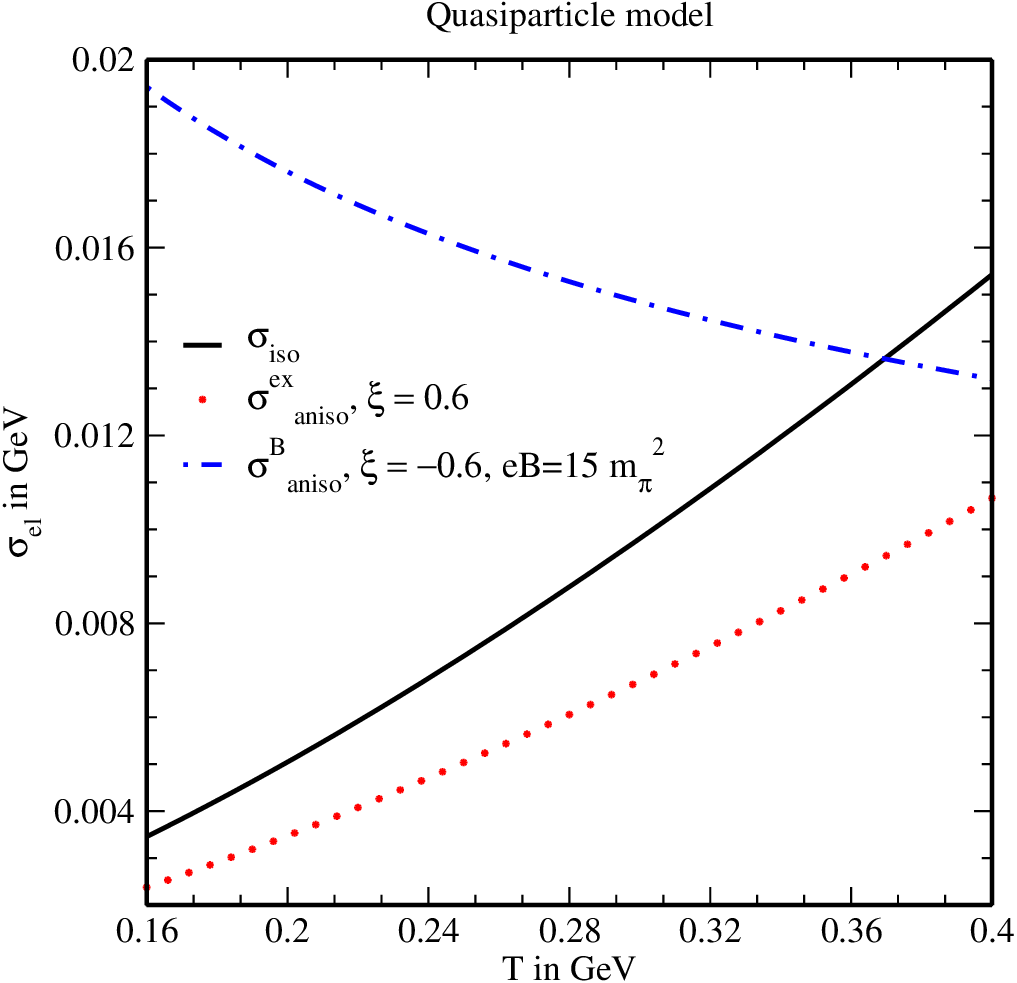}
\caption{Variation of electrical conductivity 
with temperature in the presence of momentum anisotropies 
both due to asymptotic expansion and strong 
magnetic field ($15$ $m_\pi^2$), where the effective quark masses have been used.}\label{el.1}
\end{center}
\end{figure}

With the quasiparticle description as input, we have now recomputed the 
electrical conductivity by substituting the temperature-dependent 
effective mass \eqref{Q.P.M.} into its expressions for the isotropic 
\eqref{I.E.C.} and expansion-driven anisotropic \eqref{A.E.C.(1)} 
mediums, and the temperature and magnetic field-dependent effective 
mass \eqref{Q.P.M.(eb)} into its expression for the magnetic 
field-driven anisotropic medium \eqref{A.E.C.(1eB)}. We have 
replotted $\ec$ as a function of temperature in 
figure \ref{el.1} and found that there is an overall 
decrease in $\ec$. Interestingly, for a magnetic 
field-driven weak-momentum anisotropy 
(denoted by dashed-dotted line), the magnitude of $\ec$ 
now becomes smaller, which is at par with its 
counterparts in isotropic and expansion-driven 
anisotropic mediums. However, $\ec$ for the magnetic 
field-driven anisotropic medium, now decreases with the 
temperature, which is opposite to its variation in the 
expansion-driven anisotropy. The above differences in the $\ec$'s can 
be understood qualitatively from the distributions seen in 
figures \ref{fu.qpm} and \ref{fup.qpm}, the relaxation time in the absence 
and presence of magnetic field, and the phase-space factor (which gets 
affected by the strong magnetic field only). We are now 
convinced that the quasiparticle description of 
particles tames the unusually large value of 
$\sigma_{\rm el}$ in the strong magnetic field.

\subsection{Thermal conductivity}
\begin{figure}[]
\begin{center}
\includegraphics[width=7.9cm]{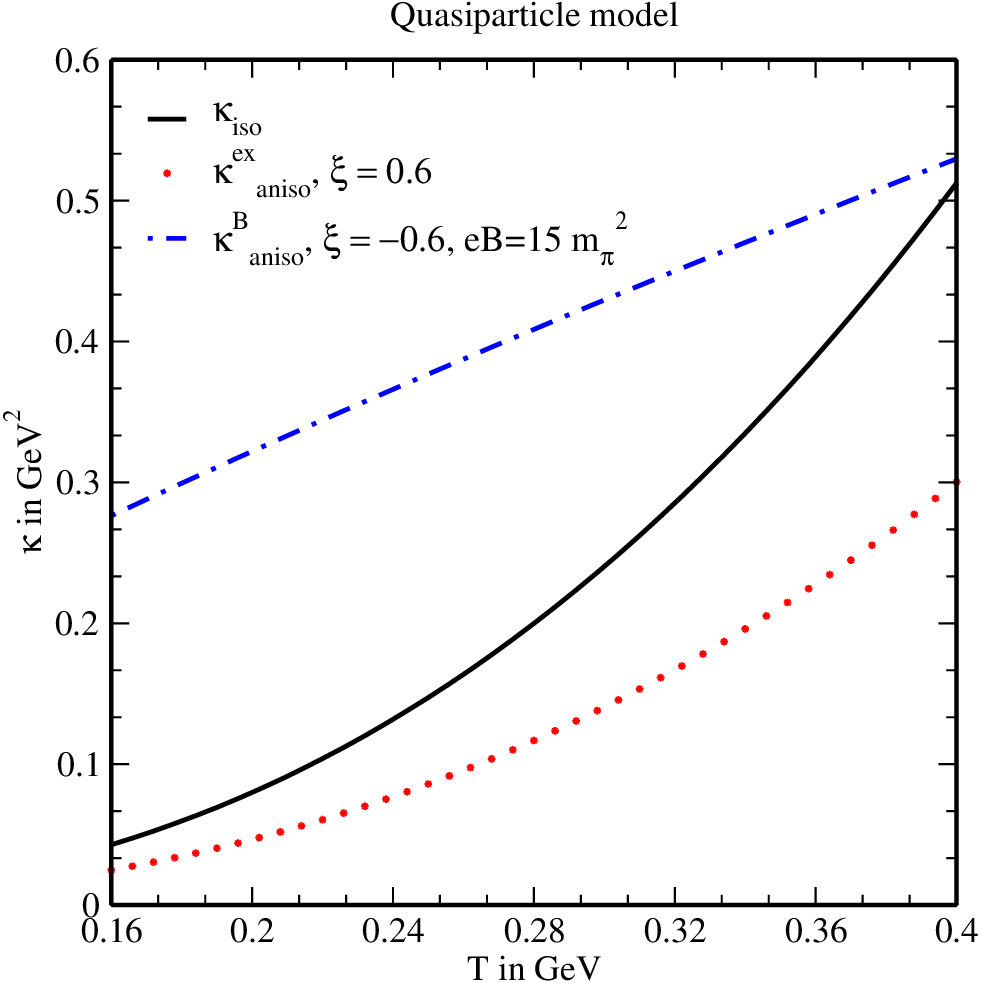}
\caption{Variation of thermal conductivity 
with temperature in the presence of momentum anisotropies 
both due to asymptotic expansion and strong 
magnetic field ($15$ $m_\pi^2$), where the effective quark masses have been used.}\label{th.1}
\end{center}
\end{figure}

We have also calculated the thermal conductivity with the 
quasiparticle description by substituting the temperature-dependent 
effective mass \eqref{Q.P.M.} into its expressions for the isotropic 
\eqref{iso.} and expansion-driven anisotropic \eqref{ex.} 
mediums, and the temperature and magnetic field-dependent effective 
mass \eqref{Q.P.M.(eb)} into its expression for the magnetic 
field-driven anisotropic medium \eqref{eb}. Figure \ref{th.1} plots the 
variation of $\kappa$ with temperature for the isotropic medium, 
expansion- and strong magnetic field-driven 
anisotropic mediums with the quasiparticle 
description. The effects of quasiparticle 
description on the thermal conductivity can again be 
understood through the distribution functions with 
quasiparticle masses in figures \ref{fu.qpm} and 
\ref{fup.qpm}, and the relaxation time in the absence 
and presence of magnetic field. For the isotropic as well as 
expansion-driven anisotropic mediums, $\kappa$ is 
found to increase with temperature as in ideal case. The 
only noticeable finding is that, although the magnitude 
of $\kappa$ for the strong magnetic field-driven anisotropic 
medium is still larger than in isotropic medium but it has now 
become smaller and comparable with the value in isotropic 
medium at higher temperature within the SMF limit ($eB \gg T^2$).

\subsection{Wiedemann-Franz law}
\begin{figure}[]
\begin{center}
\includegraphics[width=7.9cm]{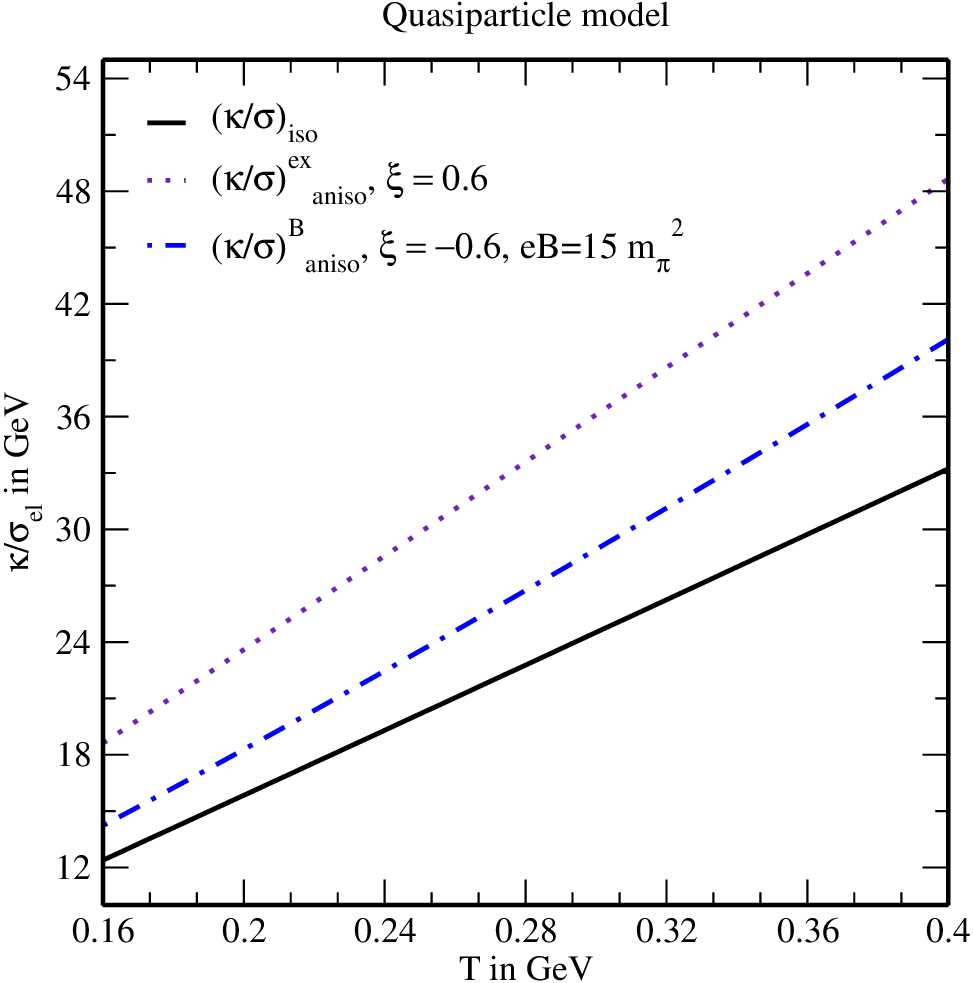}
\caption{Variation of the ratio of thermal conductivity to 
electrical conductivity with temperature in the presence 
of momentum anisotropies both due to asymptotic 
expansion and strong magnetic field ($15$ $m_\pi^2$), where the effective 
quark masses have been used.}\label{wfl}
\end{center}
\end{figure}

Wiedemann-Franz law makes us understand the relation 
between the charge transport and the heat transport 
in a system. Here we have revisited the law in 
quasiparticle description of particles, unlike the 
ideal description of particles earlier in 
previous subsection 4.1. In figure \ref{wfl} we 
found that the ratio, $\kappa/\sigma_{\rm el}$ in 
magnetic field-driven anisotropy increases linearly 
with the temperature, with a magnitude larger than that in 
isotropic medium and smaller than that in 
expansion-driven anisotropic medium. So, it can be 
used to distinguish the anisotropies of different 
origins. Thus, the Lorenz number, defined as the slope 
of the ratio ($\kappa/\ec$) versus $T$ graph, is 
smaller in the strong magnetic field-driven anisotropic medium as 
compared to its value in the expansion-driven anisotropic medium.

\subsection{Knudsen number}
\begin{figure}[]
\begin{center}
\includegraphics[width=7.9cm]{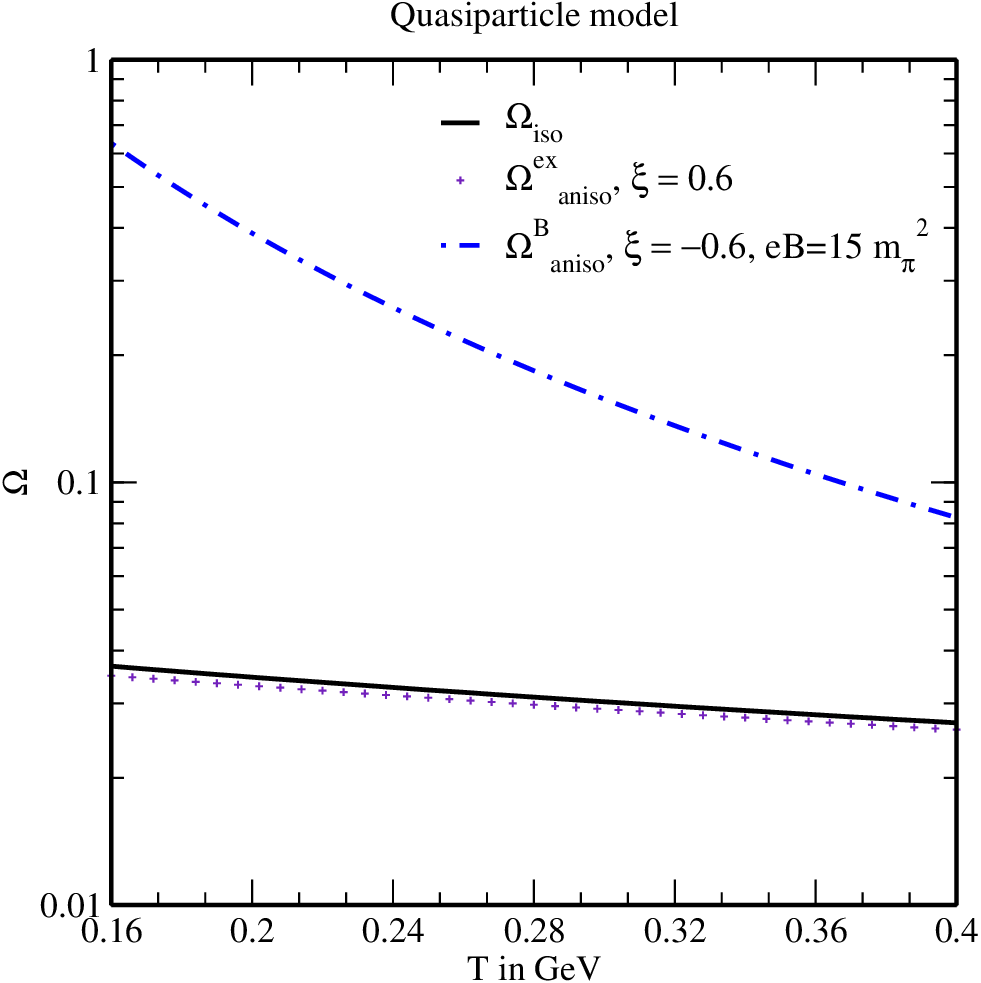}
\caption{Variation of the Knudsen number 
with temperature in the presence of momentum anisotropies 
both due to asymptotic expansion and strong 
magnetic field ($15$ $m_\pi^2$), where the effective 
quark masses have been used.}\label{frac.aniso}
\end{center}
\end{figure}

We have seen earlier that for a strong magnetic field-driven 
anisotropic medium, the Knudsen number ($\Omega$) in the ideal 
case (seen in figure \ref{frac.ideal}) was very large. As a 
result, the thermal medium in the presence of strong magnetic 
field deviates much away from its equilibrium which is, 
however, not desirable. This is exactly circumvented here in the 
quasiparticle description in figure \ref{frac.aniso}, where 
we have found that $\Omega$ has now been decreased 
drastically in the presence of strong magnetic field at par 
with the estimates for $B=0$ cases. However, there is an 
overall decrease of Knudsen number for all cases. Thus in the 
quasiparticle description, the probability of finding the 
system to be in local equilibrium is higher, due to the 
smaller value of Knudsen number.

\section{Conclusions and future outlook}
In this work, we have studied the effect of strong magnetic 
field-driven anisotropy on the transport coefficients such 
as electrical conductivity and thermal conductivity of the 
hot QCD matter and compared them with their behavior in the 
expansion-driven anisotropy. In order to find these 
conductivities we have solved the relativistic Boltzmann transport 
equation in relaxation-time approximation, where the interactions 
are incorporated through the distribution function within the 
quasiparticle approach at finite temperature and strong magnetic field. 
We have also compared the conductivities with their corresponding 
values in the ideal scenario. 

First we have revisited the formulation of 
electrical and thermal conductivities for the isotropic thermal 
medium and then calculated these for the expansion-induced 
anisotropic thermal medium. Using the value of electrical 
conductivity we have then observed the variation of 
magnetic field with time and this explains that the lifetime 
of the strong magnetic field becomes larger for an electrically 
conducting medium as compared to the vacuum, hence the strong 
magnetic field is expected to affect the charge transport and the 
heat transport in the QCD medium and this motivated us to 
derive the aforesaid conductivities for a thermal medium in the 
presence of strong magnetic field-induced anisotropy. We have 
observed that both the electrical and thermal conductivities 
have larger values in the presence of strong magnetic 
field-driven anisotropy as compared to their respective values 
in the isotropic medium, however if the anisotropy is induced 
due to asymptotic expansion, then the values of the 
conductivities are seen to get lowered than their 
values in the isotropic medium. So, in the two different types 
of anisotropic mediums, we noticed nearly opposite behavior of 
conductivities. The noticeably large values of conductivities in a strong 
magnetic field in case of ideal description are avoided using the 
quasiparticle description. Next, we have studied the Wiedemann-Franz 
law to see the relative behavior of electrical conductivity and thermal 
conductivity, where their ratio ($\kappa/\ec$) is found to increase 
linearly with temperature, but with a magnitude larger than in isotropic 
medium and smaller than in expansion-driven anisotropic medium, thus it 
can be used as a promising tool to probe the anisotropies of different 
sources. Then, we have calculated the Knudsen number to observe whether the 
system is still in equilibrium in the presence of weak-momentum 
anisotropy which may be caused by either sources. We have found that, in the 
quasiparticle description, the Knudsen number becomes less than one, thus the 
medium may remain in local equilibrium even in the presence of 
weak-momentum anisotropy.

In summary, the anisotropy affects the electrical and thermal
conductivities, which in turn affects the Lorenz number, Knudsen number
significantly. Thus it becomes imperative to suggest on the possible
signatures of the abovementioned observations in heavy-ion phenomenology
as our future plan. Chiral magnetic effect could be one such observable effect, 
because this effect is associated with the 
generation of current along the direction of magnetic 
field, which in turn depends on the magnitude of the electrical 
conductivity \cite{Fukushima:PRD78'2008}.
On the other hand the thermal conductivity could be used as a handle to 
decipher the assumption of local equilibrium in terms of mean free path via the 
inclusive production of dilepton and photon. It has been observed previously 
that the dilepton and photon yields 
get enhanced in the presence of weak-momentum anisotropy due to
initial asymptotic expansion \cite{Martinez:PRC78'2008,Ryblewski:PRD92'2015}. 
Therefore, a similar study in the 
dilepton and photon productions due to the magnetic field-driven 
anisotropy needs to be done and the comparison with the former anisotropy 
could be an indicator of the findings in thermal conductivity.

\section{Acknowledgment}
One of us (B. K. Patra) is thankful to Council of Scientific and 
Industrial Research (No.: 03(1407)/17/EMR-II), 
Government of India for the financial support of this work.

\end{document}